\documentclass[10pt,twocolumn,oneside]{IEEEtran}

\usepackage{amsmath,amsfonts,amssymb,bm}
\usepackage[tight,footnotesize]{subfigure}
\usepackage{graphicx,stfloats}
\usepackage{algorithm}
\usepackage{algorithmic}
\usepackage{cite,url}
\usepackage{cases}
\usepackage{theorem} 
\usepackage{psfrag}

\allowdisplaybreaks[4]  
\newtheorem{Theorem}{Theorem}
\newtheorem{Proposition}{Proposition}

\newtheorem{Corollary}{Corollary}[Proposition]
\theorembodyfont{\rmfamily}

\newcommand{\Cset}{\ensuremath{\mathbb{C}}}
\newcommand{\Eset}{\ensuremath{\mathbb{E}}}
\IEEEoverridecommandlockouts    
\makeatletter

\newcommand{\Rmnum}[1]{\expandafter\@slowromancap\romannumeral #1@}
\makeatother
\newcommand\numberthis{\addtocounter{equation}{1}\tag{\theequation}}

\begin{document}

\title{Capacity Analysis for Spatially Non-wide Sense Stationary Uplink Massive MIMO Systems}

\author{Xueru~Li$^\star$,~\IEEEmembership{} Shidong~Zhou$^\star$,~\IEEEmembership{\normalsize Member,~IEEE,} Emil~Bj{\"o}rnson$^{\dagger}$,~\IEEEmembership{\normalsize Member,~IEEE,} and~Jing~Wang$^\star$,~\IEEEmembership{\normalsize Member,~IEEE}
\thanks{Manuscript created March 12, 2015; revised June 15, 2015; accepted July 26, 2015. This work is supported by National Basic Research Program (2012CB316000), National Natural Science Foundation of China (61201192), National High Technology Research and Development Program of China (2014AA012107), National S\&T Major Project (2014ZX03003003-002) and Tsinghua-Qualcomm Joint Research Program. The associate editor coordinating the review of this paper and approving it for publication was J. Zhang.}
\thanks{$^\star$X.~Li, S.~Zhou and J.~Wang are with the Department of Electronic Engineering, Research Institute of Information Technology, Tsinghua National Laboratory for Information Science and Technology (TNList), Tsinghua University, Beijing 100084, China (e-mail: xueruli1206@163.com; zhousd@mail.tsinghua.edu.cn; wangj@mail.tsinghua.edu.cn).}
\thanks{$^\dagger$E.~Bj{\"o}rnson is with the Department of Electrical Engineering (ISY), Link{\"o}ping University, SE-58183 Link{\"o}ping, Sweden (e-mail: emil.bjornson@liu.se).}
\thanks{Digital Object Identifier 10.1109/TWC.2015.2464219}}

\markboth{ACCEPTED BY IEEE TRANSACTIONS ON WIRELESS COMMUNICATIONS, JUL. 2015}%
{ACCEPTED BY IEEE TRANSACTIONS ON WIRELESS COMMUNICATIONS, JUL. 2015}

\maketitle

\begin{abstract}
Channel measurements show that significant spatially non-wide-sense-stationary characteristics rise in massive MIMO channels. Notable parameter variations are experienced along the base station array, such as the average received energy at each antenna, and the directions of arrival of signals impinging on different parts of the array.
In this paper, a new channel model is proposed to describe this spatial non-stationarity in massive MIMO channels by incorporating the concepts of partially visible clusters and wholly visible clusters. Furthermore, a closed-form expression of an upper bound on the ergodic sum capacity is derived for the new model, and the influence of the spatial non-stationarity on the sum capacity is analyzed. Analysis shows that for non-identically-and-independent-distributed (i.i.d.) Rayleigh fading channels, the non-stationarity benefits the sum capacity by bringing a more even spread of channel eigenvalues. Specifically, more partially visible clusters, smaller cluster visibility regions and a larger antenna array can all help to yield a well-conditioned channel, and benefit the sum capacity. This shows the advantage of using a large antenna array in a non-i.i.d. channel: the sum capacity benefits not only from a higher array gain, but also from a more spatially non-stationary channel. Numerical results demonstrate our analysis and the tightness of the upper bound.
\end{abstract}

\begin{IEEEkeywords}
Massive MIMO, spatially non-wide sense stationary channel, upper bound on sum capacity.
\end{IEEEkeywords}
\section{Introduction} \label{sec:intro}
Massive multiple-input-multiple-output (MIMO) is a technology that has drawn considerable interest in recent years. The idea is to employ a very large number (tens or hundreds) of antennas at a base station (BS) and to serve tens of users in the same time-frequency slot~\cite{Marzetta10}. Very high spectral efficiency can be achieved without using extra spectral resources~\cite{Marzetta10, Rusek13, Larsson14,bjornson2014massive}. Intra-cell interference and thermal noise can be averaged out using simple linear signal processing like zero forcing (ZF)~\cite{Marzetta10, Yang13, Zhang13}. Thus, low complexity linear processing can be applied to achieve performance close to non-linear capacity-achieving processing such as successive interference cancellation (SIC). In addition, huge improvement in system energy efficiency can also be achieved in massive MIMO systems~\cite{Hien13, bjornson2014optimal}. The reduced transmit power per antenna further enables us to employ inexpensive components of lower quality~\cite{Larsson14, bjornson2014nonideal}. These attractive features make massive MIMO one of the key technologies for the next generation wireless communication networks.

Existing investigations on massive MIMO are mostly based on conventional channel models that are suitable for ``standard'' MIMO systems, such as the i.i.d. Rayleigh fading model in~\cite{Marzetta10, Yang13, Hien13, Jose11, Zheng2014, Wu2014}, the Kronecker model in~\cite{Shin2003} and the double-scattering model in~\cite{Hoydis2011asymptotic}. However, these models can not illustrate the non-wide sense stationary (non-WSS) characteristics of the practical massive MIMO channels. For example, measurements for massive MIMO channels in~\cite{payami2012} show that the statistical properties of the received signal vary significantly over the large array, such as the average received power and the signal directions of arrivals (DOAs). 
Denote the uplink channel from a single-antenna user to the $k$-th antenna at BS at a coherence block as $h_k$, then $h_k$ is a discrete complex random sequence in the spatial domain (i.e., with respect to the antenna index $k$). According to the definition of WSS complex random sequence in Chapter 7 of~\cite{Veerarajan2003}, $h_k$ is WSS if the expectation $\Eset \left[ h_k \right]$ is a constant that is independent of $k$, and the covariance $\rho_{kl} = \Eset[{h_k^{*} h_l }]$ depends solely on $k-l$, instead of the values of $k$ and $l$. However, the observations in~\cite{payami2012} indicate the dependence of $\rho_{kl}$ on $k$ and $l$ in massive MIMO channels. Consequently, massive MIMO channels cannot be regarded as WSS in the spatial dimension, and therefore, cannot be described directly by the aforementioned channel models.
\setcounter{footnote}{1}
\footnotetext{For a linear array of $M$ antennas with the antenna spacing being $d$, the Rayleigh distance is defined as $2d^2M^2/\lambda$ where $\lambda$ is the wavelength~\cite{Yaghjian1986}.}
\setcounter{footnote}{0}

\setcounter{figure}{0}
\begin{figure*}[t]
    \centering
    \subfigure[cluster visibility at user side.]{
    \scalebox{0.46}{\includegraphics{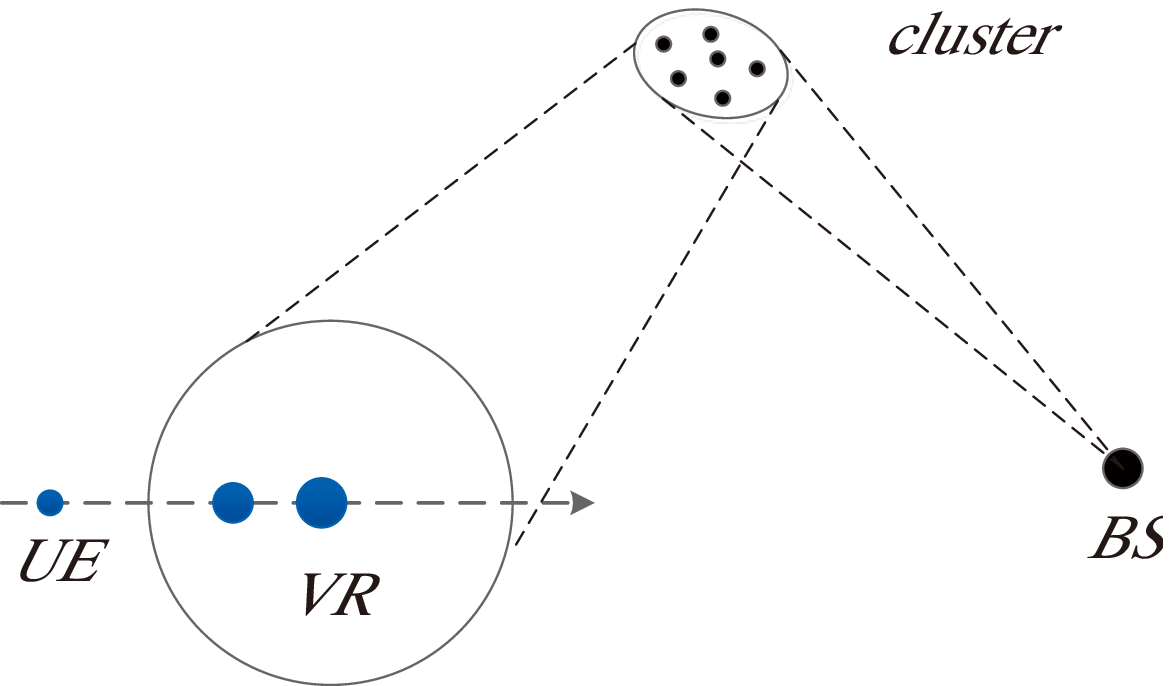}} \label{fig:VR-UE}} 
    \hspace{1cm}
    \subfigure[cluster visibility at BS side.]{
    \scalebox{0.46}{\includegraphics*{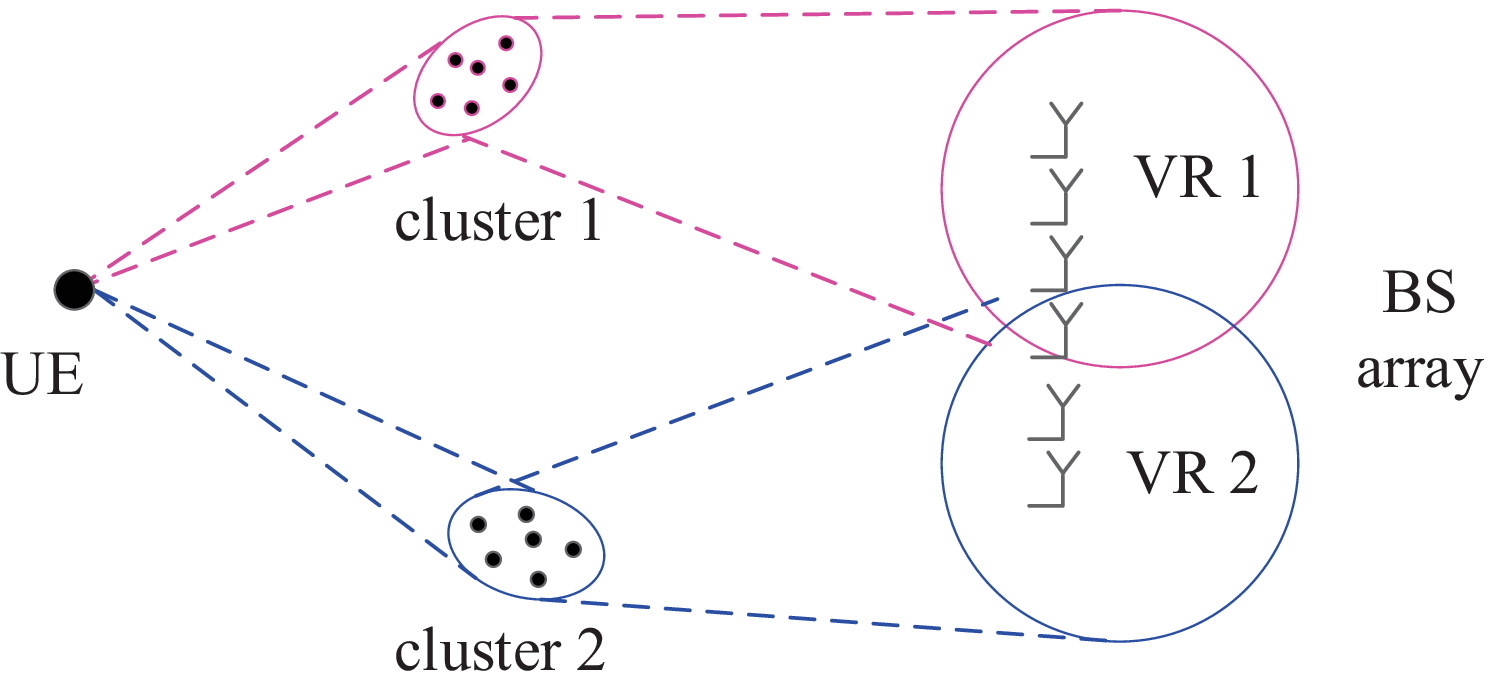}} \label{fig:VR-BS}} 
    \caption{Illustration of the cluster visibility, at (a) user side and (b) BS side. } \label{fig: VR}
\end{figure*}

Based on the analysis in~\cite{Gao13, Gao14, Wu2014channel}, the non-stationarity of massive MIMO channels results from the following two aspects. First, since the distances between the BS antenna array and some clusters are smaller than the Rayleigh distance when the number of antennas is large, the far-field propagation assumption is no longer valid.\footnotemark{ }Consequently, the wavefront should be modeled as spherical wavefront instead of plane, which brings the shift of DOAs of signals along the array. In addition, many clusters are only visible to a part of the large array due to their directions, sizes, shapes and obstacles between them and the array. Thus, two more-separated antennas are less likely to share the same set of clusters, which also leads to the variations of parameter on array axis, such as DOAs of signals and average received power. According to the measurements in~\cite{payami2012}, the first aspect is obvious for line-of-sight (LOS) users, while the second is significant for non-line-of-sight (NLOS) users.\footnotemark{ }%

\footnotetext{A LOS user is a user who has a LOS propagation path between itself and the BS, and a NLOS user is a user who do not have a LOS propagation path.}

To capture the spatially non-WSS characteristics of massive MIMO channels, the authors of~\cite{Gao13, Wu2014channel} have done some pioneering modelling work. The concept of visibility region (VR) in the COST 2100 model was employed in~\cite{Gao13}. A one-dimensional VR was modeled at the BS side and a cumulative distribution function (CDF) of the VR size was proposed. In~\cite{Wu2014channel}, a theoretical 3-D non-stationary wide band channel model for massive MIMO was proposed. It includes a birth-death process to describe the appearance and disappearance of a cluster on the array axis and a spherical wavefront assumption. By investigating their impacts on the statistical channel properties, the authors show that the spatially WSS assumption is not valid for massive MIMO channels. These geometrical models in~\cite{Gao13} and~\cite{Wu2014channel} are suitable for system level simulations. However, they are too complex for theoretical analysis of performance metrics such as channel capacity.

In this paper, we propose a new tractable massive MIMO channel model to describe the non-WSS channel characteristics in the spatial domain. Specifically, clusters are divided into two categories: wholly visible (WV) clusters that are visible to the entire array (e.g., very large buildings), and partially visible (PV) clusters that are visible only to a part of the array (e.g., small buildings, trees and cars, etc.). Then the channel spatial non-stationarity can be modeled by incorporating the two kinds of clusters and the corresponding parameters. 
Based on the new model, a closed-form expression of an upper bound on the ergodic sum capacity is further derived, and the influence of the spatial non-stationarity is analyzed. Analysis shows that for non-i.i.d. Rayleigh channels, the non-stationarity from the cluster partial visibility can increases the sum capacity by bringing a more even spread of channel eigenvalues. This shows the advantage of using a large antenna array in a practical non-i.i.d. Rayleigh fading channel: the sum capacity will benefit not only from a higher array gain, but also a more spatially non-WSS   channel. Numerical results demonstrate our analysis and the tightness of the upper bound.

The rest of this paper is organized as follows. Section~\ref{sec:pre} gives a concise introduction of the VR in the COST 2100 model. The new channel model is proposed in Section~\ref{sec:syst}. An upper bound on the ergodic sum capacity is derived in Section~\ref{sec:up-bound}. Simulation results are provided in Section~\ref{sec:simulation} before we conclude the paper in Section~\ref{sec:conclusion}. All proofs are deferred to the appendix.

\textit{Notations}: Boldface lower and upper case symbols represent vectors and matrices, respectively. The trace, transpose, conjugate, Hermitian transpose and matrix inverse operators are denoted by $\rm{tr}(\cdot )$, $(\cdot )^T$, $(\cdot )^{*}$, $(\cdot )^H$ and $(\cdot )^{-1}$, respectively.

\section{Preliminaries on the COST 2011 model} \label{sec:pre}
The COST 2100 MIMO channel model is a geometry-based stochastic channel model that was built upon the framework of the earlier COST 259 and COST 273 models~\cite{correia2006mobile,liu2012cost}. It can reproduce the stochastic properties of MIMO channels over time, frequency and space. An important concept is the VR of a scattering cluster. Given BS location, the VR of a cluster is where users can receive energy scattered from that cluster. It is a circular region of fixed size. As shown in Fig.~\ref{fig:VR-UE}, after the terminal moves into the VR, it receives signals scattered by the related cluster, and as it moves towards the VR center, the cluster smoothly increases its visibility. This visibility is accounted for mathematically by a VR gain, which grows from 0 to 1 upon entrance of the VR. By employing the VR and the VR gain, terminal mobility is supported in the COST 2100 model.

These two concepts can also be extended to the BS side, especially in massive MIMO systems, to illustrate the spatially non-WSS stationary channel characteristics. As shown in Fig.~\ref{fig:VR-BS}, different BS antennas can be within the VRs of different clusters. If the two clusters provide unequal energy and DOAs, then the received energy and DOAs of the signals vary at each BS antenna, and the channel is non-WSS in spatial dimension. From this perspective, the channel model can be built on a cluster basis.
\setcounter{figure}{1}
\begin{figure*}[ht]
\centering
\scalebox{0.46}{\includegraphics*{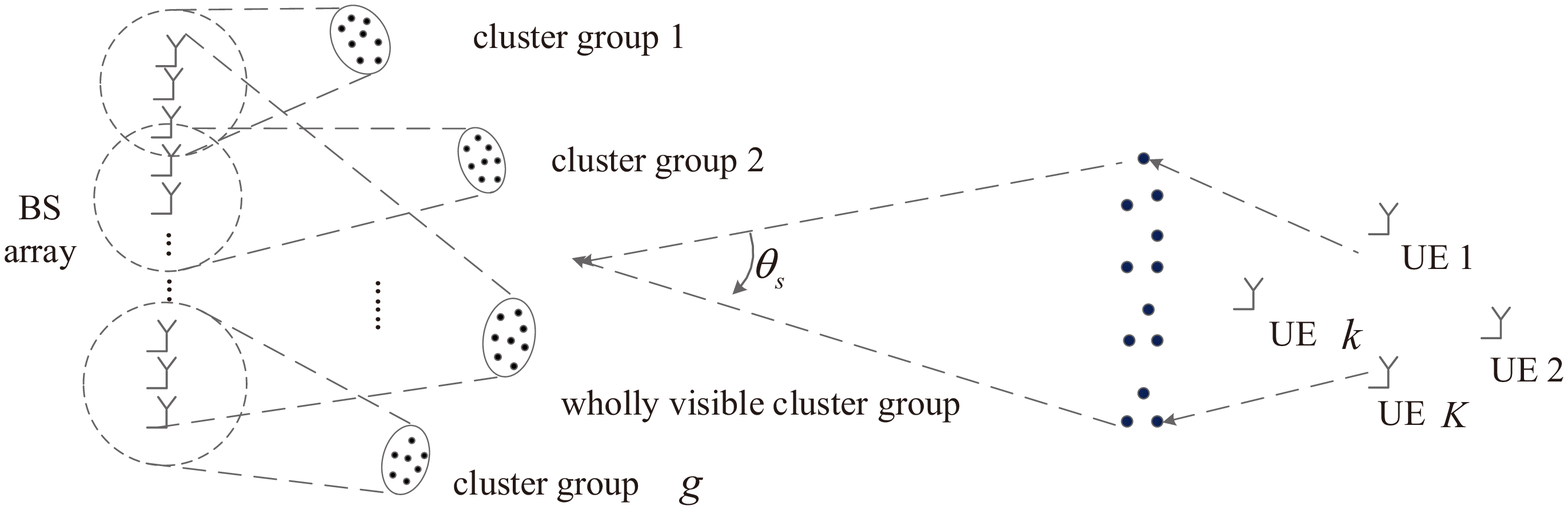}} 
\caption{Illustration of the uplink propagation scenario.}
\label{fig:uplink}
\end{figure*}

\section{Spatially Non-stationary Channel Model} \label{sec:syst}
\subsection{Scenario description} \label{subsec:description}
We consider a single-cell multi-access uplink massive MIMO system. The BS is equipped with a linear array of $M$ antennas, with the antenna spacing being ${d_r}$. $K$ single-antenna NLOS users are served simultaneously in the same time-frequency slot, and they are assumed to fall into the same geographical group, i.e., observe the same transmitter-side local clusters, as shown in Fig.~\ref{fig:uplink}. Although in the same group, the users are assumed to be spatially separated by at least a few wavelengths, so the channels of different users are mutually independent. Although it is suggested in~\cite{Huh2012, Yin2013} that users from different geographical groups should be served to suppress inter-user interference, we focus on the same-group scenario since more than one users in each group will be served when the number of users is large.

The propagation paths between each transmitter and the receiver are obstructed on both sides of the link by a set of scattering clusters. Signals emitted by the users first impinge on the transmitter-side local clusters (such as buildings, trees and cars), and then arrive at $S$ receiver-side clusters (R-clusters).\footnotemark{ }The DOAs of signals that arrive at the R-clusters span an angular spread of ${\theta _s}$. After being captured by the R-clusters, the radio signals are further reradiated to the BS. The R-clusters can be viewed as an array of $S$ virtual antennas with inter-antenna spacing $d_s$. As mentioned in the introduction, for NLOS users, many clusters are only visible to a part of the large antenna array at the BS, which is an important factor for the spatially non-WSS channel characteristics. Therefore, we divide the $S$ clusters into two categories: $s_w$ WV clusters that are visible to the whole array, and $s_p=S- s_w$ PV clusters that are only visible to a part of the array. It is also shown by~\cite{payami2012} that the VRs of many clusters cover almost the same antennas. Therefore, we further divide the $s_p$ PV clusters into $g$ groups according to their VRs, and assume that the VRs of clusters in the same group cover the same antennas at the BS. It is also shown by~\cite{payami2012} that the VRs of different groups may cover a few the same antennas. But to facilitate our analysis, we assume that the VRs of different PV groups do not overlap. For signals reradiated from the WV clusters, their DOAs at the BS antenna array span an angular spread of $\theta_w$. For signals coming from the $i$th PV cluster group which consists of ${s_{p,i}}$ clusters and covers $r_{p,i}$ consecutive antennas at the BS, the DOAs at the corresponding antenna sub-array span an angular spread of $\theta_{p,i}$. Then $\sum\nolimits_{i = 1}^{{g}} {{s_{p,i}}}  = s_p$ and $\sum\nolimits_{i = 1}^{{g}} {{r_{p,i}}} = M$. Without loss of generality, it is assumed that the first $s_p$ clusters are PV clusters, the last $s_w$ clusters are WV clusters, and the indices of antennas covered by the $i$th PV group are $\{1+\sum\nolimits_{m=1}^{i-1} r_{p,i},\ldots,\sum\nolimits_{m=1}^i r_{p,i}\}$. We further assume that the channels from the PV cluster groups to their sub-arrays, and the channel from the WV clusters to the whole array, are spatially WSS.
\footnotetext{Each cluster consists of several scatterers which are modeled as omnidirectional ideal reflectors.}

{\textbf{Remarks}}: The VRs of different PV groups do not overlap is an ideal case to some degree. But the investigation of this special case enables us to shed some light on the influence of the spatial non-stationarity on channel capacity, since channels in this case have more clear structures. Moreover, the WSS massive MIMO channel studied in most existing literature is actually another ideal case where the VRs of all clusters cover the whole antenna array. The study of the two cases can lay a foundation for the analysis of general scenarios.

\subsection{Channel modeling} \label{subsec:model}
Based on the propagation described in Subsection~\ref{subsec:description}, the multi-access channel can be modelled as
\begin{eqnarray}\label{newmodel}
{\bf{G}} = \left[ {\begin{array}{*{20}{c}}
{\bf D}_p^{\frac{1}{2}} {\bf R}_p^{\frac{1}{2}}{\bf H}_p &{{\sqrt {\frac{\rho _w}{s_w}} {\bf{R}}_w^{\frac{1}{2}}{\bf{H}}_w}}
\end{array}} \right]{\bf{R}}_s^{\frac{1}{2}}{\bf{H}} \in \mathbb{C}^{M \times K},
\end{eqnarray}
where ${\bf D}_p ={\rm{bdiag}}\{\frac{\rho_{p,1}}{{s_{p,1}}}{\bf I}_{r_{p,1}},\ldots,\frac{\rho_{p,g}}{{s_{p,g}}}{\bf I}_{r_{p,g}}\} \in \Cset^{M \times M}$, ${\bf R}_p={\rm{bdiag}}\{ {\bf{R}}_{p,1},\ldots, {\bf{R}}_{p,g}\} \in \Cset^{M \times M}$, and ${\bf H}_p = {\rm{bdiag}}\{{\bf{H}}_{p,1},\ldots,{\bf{H}}_{p,g}\} \in \Cset^{M \times s_p}$. The function ${\rm{bdiag}}\{{\bf X}_1,\ldots,{\bf X}_N\}$ creates a block-diagonal matrix with ${\bf X}_i$ being its $i$th diagonal block. ${{\bf{R}}_s} = {\bf R}_s (d_s, \theta_s)\in \Cset^{S \times S}$ is the receiver spatial correlation matrix yielded by the propagation from the users to the virtual antenna array. ${\bf R}_{p,i} = {{\bf{R}}_{p,i}}( {d_r},\theta _{p,i}) \in {\mathbb{C}^{{r_{p,i}} \times {r_{p,i}}}}$ represents the receiver correlation matrix resulted from the propagation from the $i$th PV group to the $i$th sub-array, and ${\rho _{p,i}} \ge 0$ is the corresponding propagation gain. ${{\bf{R}}_w} ={\bf R}_w ( d_r,\theta _w)\in {\mathbb{C}^{M \times M}}$ and $\rho_w \ge 0$ are similar parameters related to the WV group. ${{\bf{H}}_{p,i}} \in {\mathbb{C}^{{r_{p,i}} \times {s_{p,i}}}}$, ${{\bf{H}}_{w}} \in {\mathbb{C}^{M \times s_w}}$ and ${\bf H} \in {\mathbb C}^{S \times K}$ consist of i.i.d. zero-mean and unit-variance Gaussian random entries, and they are mutually independent.

{\textbf{Remark 1}}: As shown above, the channel spatial correlation matrix is governed by the angular spread and the antenna spacing. Take ${\bf R}_s$ as an example. It can be calculated based on the geometrical channel model as~\cite{Gesbert02}
\begin{equation}
{\bf R}_s = \sum\limits_{p=1}^P \alpha_p \Eset\left[{\bf a}\left( \beta_p\right) {\bf a}\left( \beta_p\right)^H\right],
\end{equation}
where $P$ is the number of propagation paths that arrive at the R-clusters and $\alpha_p$ is the attenuation of the $p$th path. DOAs $\beta_p$ span an angular spread $\theta_s$ and
\begin{equation}
 {\bf a}(\beta_p)=\left[1,e^{\left(-j\Delta\cos(\beta_p)\right)},\ldots,e^{\left(-j\Delta(S-1)\cos(\beta_p)\right)}\right]^T
\end{equation}
is the steering vector corresponding to $\beta_p$, where $\Delta = 2\pi d_s/\lambda$. Different assumptions on the statistics of $\beta_p$ yield different expressions for ${\bf R}_s$~\cite{Ertel1998, Fuhl1998}. In general, a larger $\theta_s$ and/or $d_s$ bring a lower correlation, and a lower $\theta_s$ and/or $d_s$ result in a higher correlation.

One can also use matrices with special structures to describe the channel spatial correlation. For example, the Toeplitz structure can reflect real-life channel statistics and cover a wide range of worst-case to best-case scenarios~\cite{Zelst2002}. Therefore, it has been widely used for many communication problems of MIMO systems~\cite{Chiani2003,Shin2006,Lozano2003,Simon2000}.

{\textbf{Remark 2}}: One may notice that our proposed model~(\ref{newmodel}) has a similar structure to the traditional double-scattering model~\cite{Gesbert02}
\begin{equation} \label{oldmodel}
{\bar {\bf G}}=\frac{1}{\sqrt{S}}{\bar {\bf R}}_r^{\frac{1}{2}}{\bar {\bf H}}_r{\bf R}_s^{\frac{1}{2}}{\bf H},
\end{equation}
where ${\bar {\bf R}}_r$ is the receiver correlation matrix observed at the BS antenna array, ${\bar {\bf H}_r}$ consists of i.i.d. Gaussian random entries and is independent of ${\bf H}$. Notice that our proposed model~(\ref{newmodel}) can reduce to model~(\ref{oldmodel}) when all clusters are WV clusters. However, as mentioned in Section I, the general case where both PV and WV clusters exist is more relevant in massive MIMO channels. In these general case, it is not straightforward to use model~(4) to investigate the influence of the spatial non-stationarity on capacity, since ${\bar {\bf R}}_r$ is influenced by both kinds of clusters. With model~(\ref{newmodel}), however, we can show clearly how the PV clusters affect the channel structure by introducing the PV-related parameters ${\bf R}_{p,i}$, $\rho_{p,i}$, and the WV-related parameters ${\bf R}_w$, $\rho_w$. Therefore, the proposition of the more tractable model~(\ref{newmodel}) is necessary for the further capacity analysis of the non-WSS massive MIMO channels.

\newcounter{TempEqCnt}
\setcounter{TempEqCnt}{\value{equation}}
\setcounter{equation}{8}
\begin{figure*}[ht]
\begin{eqnarray}\label{capacity:new}
{C_{up}}& = {\log _2}\left( {\sum\limits_{k = 0}^{\min \left\{ {M,K,S} \right\}} {{{\left( \mu  \right)}^k}\left( {k!} \right)C^{k}_{K}\sum\limits_{1 \le {i_1} \le \ldots \le {i_k} \le {S}} {\det \left( {\left( {{\bf{R}}_s} \right)_{{i_1},\ldots,{i_k}}^{{i_1},\ldots,{i_k}}} \right)} } } \right.  \nonumber \\
&\times \left. {\sum\limits_{1 \le {j_1} \le \ldots \le {j_k} \le 2M} {\det \left( {\left( {\bf{R}}_r \right)_{{j_1},\ldots,{j_k}}^{{j_1},\ldots,{j_k}}} \right)} N\left( {{j_1},\ldots,{j_k},{i_1},\ldots,{i_k}} \right)} \right)
\end{eqnarray}
\hrule
\end{figure*}
\setcounter{equation}{\value{TempEqCnt}}

Based on~(\ref{newmodel}), the channel receiver correlation matrix is
\begin{equation} \label{eqn:correlation}
\Eset\left[ {{\bf{G}}{{\bf{G}}^H}} \right] = K\left( {\bf \Lambda}_p {\bf D}_p {\bf R}_p + {\rm{tr}}\left( {{{\bf{\Lambda }}_w}} \right)\frac{{{\rho _w}}}{{{s_w}}}{{\bf{R}}_w} \right),
\end{equation}
where ${\bf \Lambda}_p= {\rm{bdiag}}\{{\rm{tr}}({\bf \Lambda}_{p,1}){\bf I}_{r_{p,1}},\ldots,{\rm{tr}}({\bf \Lambda}_{p,g}){\bf I}_{r_{p,g}}\}$. The diagonal matrix ${\bf{\Lambda}}={\rm{bdiag}} \left({\bf \Lambda}_{p,1},...,{\bf \Lambda}_{p,g},{\bf \Lambda}_w \right)\in \Cset^{S \times S}$ contains the eigenvalues of ${{\bf{R}}_s}$, with ${\bf \Lambda}_{p,i} \in \Cset^{s_{p,i} \times s_{p,i}} $ and ${\bf \Lambda}_w \in \Cset^{s_w \times s_w}$ containing the eigenvalues related to the $i$th PV cluster group and the WV group, respectively. In ${\bf R}_g$, since the channel from each cluster group to the corresponding sub-array is WSS by assumption, the diagonal elements of ${{\bf{R}}_{p,i}}$ are the same, and they can be incorporated into $\rho_{p,i}$. Thus, the diagonal elements of ${{\bf{R}}_{p,i}}$ can be reduced to unity. Furthermore, ${\rm{tr}}({\bf \Lambda}_{p,i})/s_{p,i}$ represents the average energy captured by a PV cluster in the $i$th group, and it can also be incorporated into $\rho_{p,i}$. 
From this point of view, $\rho_{p,i}$ represents the large-scale propagation gain from a user to the $i$th sub-array through a cluster of the $i$th PV group. With similar analysis for ${\bf R}_w$ and ${\rm{tr}}({\bf \Lambda}_w)/s_w$, we can consider $\rho_w$ as the propagation gain from a user to a BS antenna through a WV cluster. Therefore, the ratio of energy contributed by the PV clusters to that contributed by the WV clusters at the $i$th sub-array is $\rho_{p,i}/\rho_w$. This ratio imposes effects on the channel structure, the correlation between channels to different BS antennas, and further on the channel sum capacity.

\section{An Upper Bound on Ergodic Sum Capacity} \label{sec:up-bound}
In this section, an upper bound on the ergodic sum capacity is derived for the proposed channel model. Perfect CSI is assumed to be known at the receiver but unknown at the transmitters.\footnotemark{ }Transmitted signal ${\bf{x}} = {\left[ {{x_1},...,{x_K}} \right]^T} \in {\Cset^{K \times 1}}$ contains independent data streams of all users, and the transmit power constraint for each user is assumed to be $\frac{P}{K}$. At each symbol interval, the received signal ${\bf{y}} \in {\Cset ^{M \times 1}}$ at the BS is
\footnotetext{The acquisition of CSI can be accomplished by uplink pilot signaling.}
\begin{equation} \label{y}
{\bf{y}} = \sqrt{P}{\bf{G}}{\bf{x}} + {\bf{n}},
\end{equation}
where $\Eset [ {{{| {{x_k}} |}^2}} ] \le \frac{1}{K}$ and ${\bf{n}} \sim \mathcal{CN} ( {0,{\sigma ^2}{\bf{I}}} )$.

By maximizing the mutual information ${\cal I}({\bf x};{\bf y}, {\bf G} )$ using the maximal transmit power, the ergodic sum capacity of this multi-access channel is given by~\cite{David2005fundamentals}
\begin{equation} \label{equ:capacity}
C = \Eset\left[ {{{\log }_2}\det \left( {{{\bf{I}}_{M}} + \mu {{\bf{G}}}{\bf{G}}^H} \right)} \right],
\end{equation}
where $\mu  = \frac{P}{{K{\sigma ^2}}}$. An exact expression of the above sum capacity has been obtained for the i.i.d. Rayleigh fading channel~\cite{Shin2003}, and several upper bounds have also been derived for the traditional double-scattering channel~(\ref{oldmodel})~\cite{Hoydis2011asymptotic,Shin2003capacity,Li2010}. However, it is difficult do derive an exact expression for our model due to the complex channel structure, and it is also not straightforward to get the corresponding upper bounds for our new model, since our new model has a different channel structure from~(\ref{oldmodel}) by introducing PV clusters. Therefore, we derive a new upper bound, and with the new upper bound, we further shed some light on the influence of the spatial non-stationarity on the channel capacity. According to the concavity of log function, the upper bound on the sum capacity can be obtained by Jensen's inequality as
\begin{equation} \label{eqn:upper_bound_2}
C \le {C_{up}} = {\log _2}\left(\Eset \left[ {\det \left( {{{\bf{I}}_{M}} + \mu {{\bf{G}}}{\bf{G}}^H} \right)} \right]\right).
\end{equation}
We will show numerically in Section~\ref{sec:simulation} that this upper bound is very tight. A closed form expression of the upper bound is given in the following subsection by following an approach proposed in~\cite{Shin2003capacity}.

\subsection{Closed form expression of the upper bound}
\begin{Theorem}  \label{theorem}
For a multi-access uplink massive MIMO channel in the form of model~(\ref{newmodel}), the upper bound in~(\ref{eqn:upper_bound_2}) on the ergodic sum capacity is given in~(\ref{capacity:new}) on top of the next page,
where $\det({\bf X})_{j_1,\dots,j_k}^{j_1,\ldots,j_k}$ is the determinant of the $k\times k$ matrix lying in the $(j_1,\ldots,j_k)$ rows and $(j_1,\ldots,j_k)$ columns of $\bf X$. ${\bf{R}}_r$ is defined as
\setcounter{equation}{9}
\begin{equation} \label{correlationR}
{\bf{R}}_r = \left[ {\begin{array}{*{20}{c}}
   {{{\bf{W}}_p}}&{{\bf{W}}_p^{\frac{1}{2}}{\bf{W}}_w^{\frac{1}{2}}}\\
   {{\bf{W}}_w^{\frac{1}{2}}{\bf{W}}_p^{\frac{1}{2}}}&{{{\bf{W}}_w}}
   \end{array}} \right] \in {\mathbb{C}^{2M \times 2M}},
\end{equation}
with ${\bf W}_p ={\bf D}_p {\bf R}_p$ and ${\bf W}_w = \frac{{{\rho _w}}}{{{s_w}}}{{\bf{R}}_w}$. $C_K^k = \frac{{K!}}{{k!\left( {K - k} \right)!}}$ is the number of k-combination of set $\left\{1,...,K \right\}$. Define a block diagonal matrix
\begin{equation}
{\bf E}={\rm {bdiag}}\left({\bf 1}_{r_{p,1}}{\bf 1}_{s_{p,1}}^T,\ldots,{\bf 1}_{r_{p,g}}{\bf 1}_{s_{p,g}}^T,{\bf 1}_M{\bf 1}_{s_w}^T\right) \in \Cset^{2M \times S},
\end{equation}
where ${\bf 1}_n =[1,\ldots,1]^T \in \Cset^{n \times 1}$. Suppose $({j_1},\ldots,{j_k})$ and $({i_1},\ldots,{i_k})$ select $m_i$ rows and $n_i$ columns from the $i$th diagonal block of $\bf E$, respectively, then $\sum\nolimits_{i=1}^{g+1}m_i=k$ and $\sum\nolimits_{i=1}^{g+1}n_i=k$ and
\begin{align*} \label{N}
   & N\left( {{j_1},...,{j_k},{i_1},...,{i_k}} \right) \nonumber \\
   & = \left\{ \begin{array}{lc}
    \prod\limits_{i = 1}^{g+1} {m_i}! & \text{if}\quad m_1=n_1,\ldots,m_{g+1}=n_{g+1}, \\
    0 & \text{otherwise}. \numberthis
    \end{array} \right.
\end{align*}
\end{Theorem}
\noindent\emph{Proof:} See Appendix~\ref{sec:proof thrm1}. \hfill{$\blacksquare$}

Theorem~\ref{theorem} shows that the influence of the channel spatial non-stationarity on the sum capacity lies in the structure of the correlation matrix ${\bf{R}}_r$ in~(\ref{correlationR}) and the value of $N$ in~(\ref{N}). For example, if all clusters are PV to the BS array, i.e., $S = s_p$ and ${\rho _w}/{s_w}=0$, then ${\bf R}_r$ reduces to the block diagonal matrix ${\bf W}_p$, which means the channels to different sub-arrays are mutually independent. Therefore, a more even spread of eigenvalues of ${\bf R}_r$ can be expected, which will bring a capacity increase. The upper bound on the sum capacity in this case is presented in Proposition~\ref{prop2}.
\begin{Proposition}\label{prop2}
If all clusters are PV clusters,  which means $S= s_p$ and $\frac{\rho _w}{s_w} = 0$, the capacity upper bound is given in~(\ref{capacity:rho=0}) on top of the next page, where ${\bf W}_p$ is defined in Theorem~\ref{theorem}.
\end{Proposition}
\setcounter{TempEqCnt}{\value{equation}}
\setcounter{equation}{12}
\begin{figure*}[ht]
\begin{eqnarray} \label{capacity:rho=0}
{C_{up}} &= {\log _2}\left( {\sum\limits_{k = 0}^{\min \left\{ {M,K,{S}} \right\}} {{\mu ^k}\left( {k!} \right)} C^{k}_{K} } \right.\left\{ {\sum\limits_{1 \le {i_1} \le \ldots \le {i_k} \le {S}} {\det \left( {\left( {{{\bf{R}}_s}} \right)_{{i_1},\ldots,{i_k}}^{{i_1},\ldots,{i_k}}} \right)} } \right. \nonumber \\
& \times \left. {\left. {\sum\limits_{1 \le {j_1} \le \ldots \le {j_k} \le M} {\det \left( {\left( {{{\bf{W}}_p}} \right)_{{j_1},\ldots,{j_k}}^{{j_1},\ldots,{j_k}}} \right) N\left( {{j_1},\ldots,{j_k},{i_1},\ldots,{i_k}} \right)} } \right\}} \right)
\end{eqnarray}
\hrule
\end{figure*}
\setcounter{equation}{\value{TempEqCnt}}

Another extreme case is that all clusters are WV to the array, i.e., $S=s_w$ and $\rho_{p,i}/s_{p,i}=0$, then ${\bf R}_r$ reduces to ${\bf W}_w$ and $N(j_1,\ldots,j_k,i_1,\ldots,i_k)=k!$, and our proposed model reduces to model~(\ref{oldmodel}). The upper bound on the ergodic sum capacity in this case can also be obtained from Theorem~\ref{theorem}, and has been studied in Theorem \Rmnum{3}.3 in~\cite{Shin2003capacity}. In a ideal rich scattering environment, our model further reduces to the i.i.d. Rayleigh fading channel by setting ${\bf W}_w =  {\bf I}_M$, ${\bf R}_s = {\bf I}_S$ and $S\to \infty$, and its capacity upper bound can also be simplified from Theorem~\ref{theorem} as
\setcounter{equation}{13}
\begin{equation}
C_{up} = \log_2\left( \sum\limits_{k=0}^{\min\{M,K\}}\left(\mu\right)^k k! C_K^k C_M^k\right),
\end{equation}
which is consistent with (22) in~\cite{Shin2003capacity}. In a practical propagation environment, however, the correlation coefficients in ${\bf W}_w$ can be very large due to the finite scattering, which will seriously compromise the sum capacity. In the following analysis, we will focus on the scenarios where spatial correlations exist.

The all-PV and all-WV examples above indicate that the number of PV clusters and the energy contributed by them could impose notable influence on the structure of ${\bf R}_r$, and then on the sum capacity. To show the influence explicitly, the influence of $\frac{\rho _{p,i}}{\rho _w}$ on the sum capacity is first investigated.

\subsection{Impact of the energy ratio $\rho_{p,i}/\rho_w$ on the sum capacity}
Before we continue, a closed form expression of the upper bound at high SNR is first obtained. Due to the factorial of $k$ and the $k$-exponent on SNR $\mu$ in~(\ref{capacity:new}), the ${\min \left\{ {M,K,S} \right\}}$th order becomes dominant in the upper bound at high SNR. If $M = S = K =n$, we have at high SNR
\begin{eqnarray}   \label{capacity:high snr}
{C_{up}} &=&  n{\log _2}\mu  + {\log _2}\left( {n!} \right) + {\log _2}\det \left( {{{\bf{R}}_s}} \right) \nonumber \\
& +& {\log _2}\left( \sum\limits_{1 \le {j_1} \le \ldots \le {j_n} \le 2n} \det \left( {\left( {{{\bf{R}}_r}} \right)_{{j_1},\ldots,{j_n}}^{{j_1},\ldots,{j_n}}} \right) \right.\nonumber \\
& \times & \left. N\left( {{j_1},\ldots,{j_n},1,\ldots,n} \right) \right).
\end{eqnarray}
The case of $M=K$ is considered since it provides the highest capacity that can be possibly obtained when $M$ is fixed. Moreover, if users are equipped with multiple antennas, then $K$ can be viewed as the number of total antennas at the transmitters, and $M=K$ can be achievable. Therefore, the analysis in the sequel will be based on~(\ref{capacity:high snr}). As we will show in Section~\ref{sec:simulation}, the conclusions for the upper bound at high SNR also hold for medium and low SNR regions, and for the setup of $M>K$. Most importantly, they hold for true capacity as well.

\begin{Proposition} \label{prop1}
Let $M = S = K =n$, $g > 1$, ${r_{p,i}} = {r_{p,0}} = \frac{M}{g}$, ${s_{p,i}} = {s_{p,0}} = \frac{s_p}{g}$, $\rho_{p,i} = \rho_p$, and $\rho_p + \rho_w =1$ (so the received energy at each BS antenna is $K$), and assume that ${\bf R}_{p,i}( i=1,...,g)$ is invertible, then the capacity upper bound in~(\ref{capacity:high snr}) is an increasing function of $\rho_p$ when $0 \le {\rho _p} \le \frac{s_p}{S}$, and a decreasing function when $\frac{s_p}{S} < {\rho _p} \le 1$. The maximum value is achieved when ${\rho _p} = \frac{s_p}{S}$ and is given by
\begin{eqnarray} \label{Cmax}
{C_{max}}& =& n{\log _2}\left( { \mu} \left( n!\right)\right) + {\log_2}\det \left( {{{\bf{R}}_s}} \right)+g{\log _2}\left( {{s_{p,0}}!} \right) \nonumber \\
&+&{\log _2}\left( {{s_w}!} \right)+{\log _2}{\left( {\frac{g}{n}} \right)^{{s_p}}} + {\log _2}{\left( {\frac{1}{{{s_w}}}} \right)^{{s_w}}} \nonumber \\
&+& {\log _2}\left(\sum\limits_{\left( {{j_1},\ldots,{j_n}} \right) \in {\cal A}} {\det \left( {\left( {{\bf{\tilde R}}_r} \right)_{{j_1},\ldots.,{j_n}}^{{j_1},\ldots,{j_n}}} \right)}\right) ,
\end{eqnarray}
where ${\bf{\tilde R}}_r $ is defined as
\begin{equation}
{\bf{\tilde R}}_r = \left[ {\begin{array}{*{20}{c}}
{\bf R}_p & {{\bf R}_p^{\frac{1}{2}}{\bf{R}}_w^{\frac{1}{2}}}\\
{{\bf{R}}_w^{\frac{1}{2}}{\bf{R}}_p^{\frac{1}{2}}}&{{{{\bf{R}}}_w}}
\end{array}} \right].
\end{equation}
$\forall \left( {{j_1},\ldots,{j_n}} \right) \in {\cal A}$, $\left( {{j_1},\ldots,{j_n}} \right)$ satisfies the condition under which $N(j_1,\ldots,j_k,1,\ldots,n)$ in~(\ref{N}) is non-zero.
\end{Proposition}
\noindent\emph{Proof:} See Appendix~\ref{sec:proof prop1}. \hfill{$\blacksquare$}

Proposition~\ref{prop1} shows that the maximum value of the upper bound is obtained when the energy proportions of the two kinds of clusters match their number proportions. The conclusion makes sense since more clusters tend to capture more energy from users and reradiate more energy to BS antennas,. 
Proposition~\ref{prop1} itself is an intermediate result which leads to a more important question: in the site selection for a massive MIMO system, if the received energy from multiple BSs are almost the same, which propagation environment is more favorable, the one with a higher number proportion of PV clusters or the one with a lower proportion? This question can be answered by studying the effect of $s_p/S$ on $C_{max}$ in~(\ref{Cmax}).

To conduct in-depth analysis, a reasonable and easy-to-use structure is needed for the correlation matrices in our model. As mentioned in Subsection~\ref{subsec:model}, the complex Toeplitz structure has advantages of reflecting real-life channel statistics and covering a wide range of scenarios. Therefore, the complex Toeplitz structure is used in the following discussion. Without loss of generality, ${\bf R}_{p,i}={\bf R}_{p,0}$ is further assumed. Then ${\bf R}_{p,0}={\bf \Omega}_{r_{p,0}}(a_p)$, ${\bf R}_w={\bf \Omega}_M(a_w)$ and ${\bf R}_s={\bf \Omega}_S(a_s)$, where ${\bf \Omega}_d(a)$ is a $d \times d$ complex Toeplitz matrix that can be written as
\begin{equation} \label{toeplitz}
{{\bf{\Omega }}_d}\left( a \right) = \left[ {\begin{array}{*{20}{c}}
1&a&a^2& \ldots &a^{d-1}\\
a^{*}&1&a& \ldots &a^{d-2}\\
(a^{*})^2&a^{*}& \ddots & \ddots & \vdots \\
 \vdots & \vdots & \ddots &1&a\\
(a^{*})^{d-1}&(a^{*})^{d-2}& \ldots &a^{*}&1
\end{array}} \right],
\end{equation}
with $0 \le |a| \le 1$.

It should be noticed that the capacity is not only influenced by the structure of ${\bf R}_r$, but also influenced by the values of the elements of ${\bf R}_r$, i.e., the values of $|a_p|$ and $|a_w|$. The relation between $|a_p|$ and $|a_w|$ (e.g., $|a_p|<|a_w|$) in a particular scenario would also influence our analysis about the impact of spatial non-stationarity on capacity. In general, however, an exact relation can not be provided, since the angular spread of a cluster group is related to the spatial distribution of the clusters belong to that group, and a wide spread of clusters can lead to a large angular spread. In practical propagation environment, a WV cluster group may consist of clusters that are co-located (e.g., a part of a wall of a building), while a PV cluster group may consist of clusters that are separated (like cars and walls of several buildings), and they are only visible to some antennas due to their directions, small sizes, or due to some obstacles between them and the antenna array. In this case, the PV group can give a larger angular spread than the WV cluster group. For the same reason, there are also cases that a WV cluster group gives a larger angular spread. In general, since the spatial distribution of clusters is not known, we assume that these two kinds of cluster group give the same (or similar) angular spread, and as a result, give the same correlation parameters, i.e., $|a_p|=|a_w|$.

For the investigation of the effect of $s_p/S$ on $C_{max}$, two special cases, $S=s_p$ and $S= s_w$, are considered here, since it is nontrivial to obtain the derivative of $C_{max}$ with respect to $s_p$. The general case with $0 < s_p < S$ is considered numerically in the next section. The following corollary is obtained.
\begin{Corollary} \label{corollary1.1}
Keeping other parameters fixed, the $C_{max}$ in~(\ref{Cmax}) achieves a higher value with $S=s_p$ than it does with $S=s_w$, i.e.,
\begin{equation} \label{Cmax compare}
{\left.C_{max} \right|_{S = s_p}} > {\left. {{C_{max}}} \right|_{S = {s_w}}},
\end{equation}
and the gap ${\left. {{C_{max}}} \right|_{S = {s_p}}}- {\left. {{C_{max}}} \right|_{S = {s_w}}}$ increases as $g$ grows.
\end{Corollary}
\noindent\emph{Proof:} See Appendix~\ref{sec:proof corollary1.1}. \hfill{$\blacksquare$}

Corollary~\ref{corollary1.1} shows that a higher capacity upper bound can be expected from a complete PV channel. As shown in Section~\ref{sec:simulation}, the behavior also holds for the true capacity. The increase of the capacity results from the fact that in a complete PV scenario, channels to different BS sub-arrays are uncorrelated, which brings a more even spread of channel eigenvalues, and then help to increase the sum capacity. Consequently, a more PV-significant environment is favorable in the site selection for massive MIMO systems, if the received energy at multiple BSs are almost the same, and the angular spreads of the WV group and the PV group are similar. This conclusion is based on the simplification assumption that $|a_p|=|a_w|$, and can extend to the cases that $|a_p|<|a_w|$ where the PV clusters contribute more to a correlation reduction. However, it may not necessarily extend to the cases that $|a_p|>|a_w|$.

\subsection{Impact of VR size and BS antenna array size}
Besides the energy ratio and the number ratio of PV clusters, the channel spatial non-stationarity is impacted by the VR sizes and the size of the BS antenna array as well. For example, if some clusters in the environment provide small VRs, more PV cluster groups are needed to cover the BS antenna array, which increases the channel non-stationarity and exerts positive effect on the sum capacity. Moreover, when a physically larger antenna array (which consists of more antennas with fixed antenna spacing) is built at the BS, it tends to see more clusters, especially PV clusters groups, which also has influence on the channel non-stationarity. Since the number of WV clusters will not increase as the array size grows, it is reasonable to assume that all clusters are PV clusters when the array size is large enough. Thus, we focus on the complete PV scenarios in this subsection.

In the following analysis, we still consider the case of $M = K = S=n$, ${r_{p,i}} = {r_{p,0}} = \frac{M}{g}$, ${s_{p,i}} = {s_{p,0}} = \frac{S}{g}$, ${\bf{R}}_{p,i}={\bf{R}}_{p,0}={{\bf{\Omega }}_{r_{p,0}}}( a ) ( i \in \{ 1,\ldots,g \})$ and ${\bf{R}}_{w}={{\bf{\Omega }}_{{M}}}( a ) ( {0 \le a \le 1} )$. Apparently, a larger $g$ brings smaller VR sizes, thus the influence of VR size can be studied by investigating the influence of $g$. The following proposition is obtained.

\begin{Proposition} \label{corollary:2.1}
Keeping all other parameters fixed, the capacity upper bound at high SNR in~(\ref{capacity:high snr}) is an increasing function of the number of PV cluster groups $g$.
\end{Proposition}
\noindent\emph{Proof:} The proof is similar to that of Corollary~\ref{corollary1.1}, thus it is omitted. \hfill{$\blacksquare$}

This result is quite intuitive. Since a larger $g$ means that the BS antenna array is divided into more sub-arrays, and channels to different sub-arrays are uncorrelated in the complete PV scenarios, a more even spread of channel eigenvalues can be expected. Therefore, Proposition~\ref{corollary:2.1} implies that placing the BS array in a more PV significant environment will bring a higher sum capacity. However, it is worth noticing that the received energy at each BS antenna is independent of the cluster number in our model (all channels to sub-arrays are normalized with respect to the cluster numbers). Thus in our analysis, although an increase of $g$ means less clusters in each cluster group, it does not result in a decrease of received energy at the BS. Consequently, Proposition~\ref{corollary:2.1} is implicitly based on the assumption that the received energy at the BS does not decrease with the increase of $g$. It is also based on the assumption that $|a_p|$ remains unchanged as $g$ grows.

To analyze the impact of array size, the number of antennas $M$, the number of clusters $S$ and the number of cluster groups $g$ are assumed to increase simultaneously. This assumption makes sense in real propagation environments, since a larger antenna array tends to see more clusters and cluster groups. Therefore we assume $n=cg$ where $c>0$ is a constant, then the increase of $n$ brings the simultaneous increase of $g$. Then for the considered scenario, the capacity upper bound at high SNR is
\begin{eqnarray}\label{c_up remark2.2}
C_{up}&=& n{\log_2}{\frac{\tilde \mu}{n}} +{\log_2}\left(n! \right) + \log_2 {\rm {det}}\left({\bf R}_s\right)\nonumber \\
&+& \frac{n}{c} \log_2 {\rm {det}}\left(\frac{1}{c}{\bf R}_{p,0} \right) + \frac{n}{c} \log_2\left(c!\right),
\end{eqnarray}
where $\tilde \mu= \frac{P}{\sigma^2}$. Applying Stirling's approximation to $n!$, and plugging the determinants of ${\bf R}_s$ and ${\bf R}_{p,0}$ into $C_{up}$, we investigate the influence of array size by investigating the monotonicity and the increasing rate of $C_{up}$ with respect to $n$. The conclusion is obtained in the following proposition.
\begin{Proposition} \label{corollary:2.2}
Keeping all other parameters fixed, $g(n)$ is an increasing function of $n$ and as $n \to \infty$,
\begin{eqnarray}
\frac{{\partial C_{up}}}{{\partial n}} \to = {\log _2}\left( {\frac{\tilde \mu }{ec}} \right) + \frac{1}{c}{\log _2}\left( {c!} \right) + \log_2 h\left(a_s, a_p\right),
\end{eqnarray}
where $h\left(a_s, a_p\right) = (1-|a_s|^2)(1-|a_p|^2)^{1-\frac{1}{c}}$.
\end{Proposition}
\noindent\emph{Proof:} See Appendix~\ref{sec:proof corollary2.2}. \hfill{$\blacksquare$}

Proposition~\ref{corollary:2.2} shows that a higher capacity can be achieved by increasing the system dimensions, and in the large system limit, i.e., $n \to \infty$, the increasing rate $\frac{{\partial C_{up}}}{{\partial n}}$ converges to a constant. Notice that we reduce the transmit power of each user by $\frac{1}{M}$ as stated in~(\ref{y}), therefore the benefit brought by a larger array gain is offset. Thus, the capacity increase comes from a higher multiplexing gain and a more non-WSS channel. Moreover, as shown in Section~\ref{sec:simulation}, the increase can still be obtained even if we fix the number of users $K$ while increasing $M$ and $S$, i.e., fix the multiplexing gain. It shows that aside from a higher array gain and a higher multiplexing gain, the sum capacity of a massive MIMO system also benefits from a more spatially non-WSS channel.

In addition, it is straightforward to prove that $\frac{{\partial C_{up}}}{{\partial n}}$ is an decreasing function of $c$, i.e., the VR size. Therefore, a complete PV scenario ($c<n$) provides a higher increasing rate of capacity than a complete WV scenario ($c=n$), which also shows the benefit of a spatially non-WSS channel.

\section{Numerical Evaluation} \label{sec:simulation}
In this section, we present numerical results to verify the tightness of the upper bound and conclusions of our analysis. In every simulation, 10000 independent Monte-Carlo realizations of channels in form of model~(\ref{newmodel}) are generated. The capacity predicted by the upper bound is compared to the numerical computation of~(\ref{equ:capacity}). The definition of SNR is $\mu = \frac{P}{K\sigma^2}$. The correlation coefficients $a_s = 0.6$, $a_w=a_p=0.85$.

\begin{figure}[H]
\centering
\scalebox{0.43}{\includegraphics*{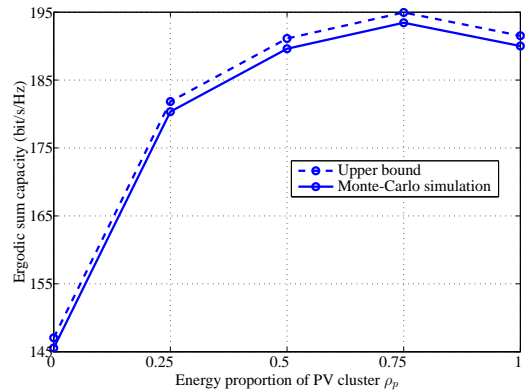}}
\caption{Ergodic sum capacity as as a function of $\rho_p$, with $M = S= 128$, $K= g =32$, $s_p=96$ and SNR=15 dB.}
\label{changerhop}
\end{figure}

The results that validate Proposition~\ref{prop1} are shown in Fig.~\ref{changerhop} with $\rho_{p,i}=\rho_p \in \left\{0, 0.25, 0.5, 0.75, 1 \right\}$. Given $M = S= 128$, $K= g =32$ and $s_p=96$, the simulation results and the analytical results are compared. As shown in the figure, the sum capacity first increases and then decreases as $\rho_p$ grows, and the maximum value is achieved when $\rho_p = \frac{s_p}{S}$. In addition, the capacity gain by increasing $\rho_p$ is notable: the capacities at $\rho_p =0.75$ and $\rho_p =1$ are $34$\% and 31\% higher than that at $\rho_p =0$, respectively. Therefore, more energy from PV clusters benefits the sum capacity. Moreover, the figure shows that the derived upper bound is very tight.
\psfrag{0}[][l]{\Large {0}}
\psfrag{0.25}[][l]{\Large {0.25}}
\psfrag{0.5}[][l]{\Large {0.5}}
\psfrag{0.75}[][l]{\Large {0.75}}
\psfrag{1}[][l]{\Large {1}}
\psfrag{145}[][]{\Large {145}}
\psfrag{155}[][]{\Large {155}}
\psfrag{165}[][]{\Large {165}}
\psfrag{175}[][]{\Large {175}}
\psfrag{185}[][]{\Large {185}}
\psfrag{195}[][]{\Large {195}}
\psfrag{rhop}[][cb]{\Large {Energy proportion of PV cluster $\rho_p$}}
\psfrag{sum capacity}[][]{\Large{Ergodic sum capacity (bit/s/Hz)}}
\psfrag{Upper bound blablablablablabla}{\Large{Upper bound}}
\psfrag{Numerical results}{\Large{Monte-Carlo simulation}}
\psfrag{5}[][l]{\Large {5}}
\psfrag{10}[][l]{\Large {10}}
\psfrag{15}[][l]{\Large {15}}
\psfrag{20}[][l]{\Large {20}}
\psfrag{25}[][l]{\Large {25}}
\psfrag{30}[][l]{\Large {30}}
\psfrag{35}[][l]{\Large {35}}
\psfrag{100}[][]{\Large {100}}
\psfrag{200}[][]{\Large {200}}
\psfrag{300}[][]{\Large {300}}
\psfrag{400}[][]{\Large {400}}
\psfrag{Indices of eigenvalues}[][cb]{\Large {Indices of eigenvalues}}
\psfrag{Eigenvalues}[][]{\Large{Eigenvalues}}
\psfrag{rhop=0 blablabla}{\Large {$\rho_p=0$}}
\psfrag{data2}{\Large {$\rho_p=0.25$}}
\psfrag{data3}{\Large {$\rho_p=0.75$}}
\psfrag{data4}{\Large {$\rho_p=1$}}
\begin{figure}[t]
\centering
\scalebox{0.42}{\includegraphics*{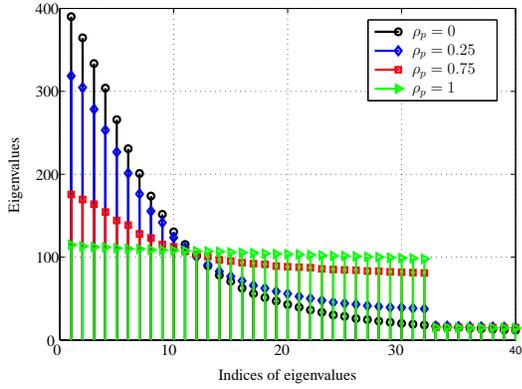}}
\caption{Eigenvalue spread as a function of $\rho_p$, with $M = S= 128$, $K= g =32$, $s_p=96$ and SNR=15 dB.}
\label{eigenvalue}
\end{figure}
\psfrag{50}[][]{\Large {50}}
\psfrag{150}[][]{\Large {150}}
\psfrag{250}[][]{\Large {250}}
\psfrag{SNR(dB)}[][cb]{\Large {SNR(dB)}}
\psfrag{Upper bound np=ns blablablablablabla}{\Large {Upper bound: $s_p=S$}}
\psfrag{data2}{\Large {Monte-Carlo simulation: $s_p=S$}}
\psfrag{Upper bound np=ns/2}{\Large{Upper bound: $s_p=\frac{S}{2}$}}
\psfrag{data4}{\Large {Monte-Carlo simulation: $s_p=\frac{S}{2}$}}
\psfrag{Upper bound np=0}{\Large {Upper bound: $s_p=0$}}
\psfrag{data6}{\Large {Monte-Carlo simulation: $s_p=0$}}

\psfrag{Upper bound g=32 aaaaaaaaaaaaaaaaaaaaaaaaa}{\Large {Upper bound: $g=32$}}
\psfrag{data22}{\Large {Monte-Carlo simulation: $g=32$}}
\psfrag{Upper bound g=16}{\Large{Upper bound: $g=16$}}
\psfrag{data42}{\Large {Monte-Carlo simulation: $g=16$}}
\psfrag{Upper bound g=1}{\Large {Upper bound: $g=1$}}
\psfrag{data62}{\Large {Monte-Carlo simulation: $g=1$}}
\begin{figure}[t]
\centering
\scalebox{0.42}{\includegraphics*{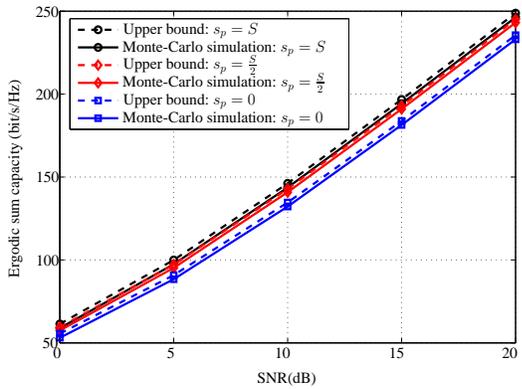}}
\caption{Maximum sum capacity as a function of $s_p$, with $M=S=128$, $K=g=32$, $s_p \in \left\{0,64,128\right\}$ and $\rho_p \in \left\{0,0.5,1\right\}$.}
\label{cmax}
\end{figure}
As analyzed before, the influence of $\rho_p$ on capacity can be perceived by showing the spreads of eigenvalues in $\Eset \left[ {{\bf{G}}{{\bf{G}}^H}} \right]$. Parameters remain the same as in Fig.~\ref{changerhop} and the results are shown in Fig.~\ref{eigenvalue}. When $\rho_p=0$ or $\rho_p=0.25$, the eigenvalue drops very fast. When $\rho_p=0.75$ or $\rho_p=1$, however, the eigenvalues show up in groups and drop very slowly within each group. This is because the eigenvalues in the same group belong to channels to different sub-arrays, and since the channels to different sub-arrays are independent, the eigenvalues in the same group do not indicate correlation and therefore, drop very slowly. Moreover, the channel to a sub-arrays is spatially correlated, thus the eigenvalues in different groups indicate correlation and drop drastically. If we focus on the first 32 dominant eigenvalues, the singularity of $\Eset \left[ {{\bf{G}}{{\bf{G}}^H}} \right]$ drops drastically as $\rho_p$ increases, which implies that notably positive influence on capacity can be brought by a larger $\rho_p$.

To validate Corollary~\ref{corollary1.1}, results of the maximal sum capacity as a function of $s_p$ is shown in Fig.~\ref{cmax}. As shown in the figure, a larger $s_p$ brings a higher sum capacity: the capacity of $s_p=S$ (i.e., a complete PV channel) is 13 bit/s/Hz higher than that of $s_p=0$ (i.e., a complete WV channel). It justifies our conclusion that the number of PV clusters, i.e., the significance of the channel non-stationarity is an important factor worth consideration in the site selection.
\begin{figure}[t]
\scalebox{0.43}{\includegraphics*{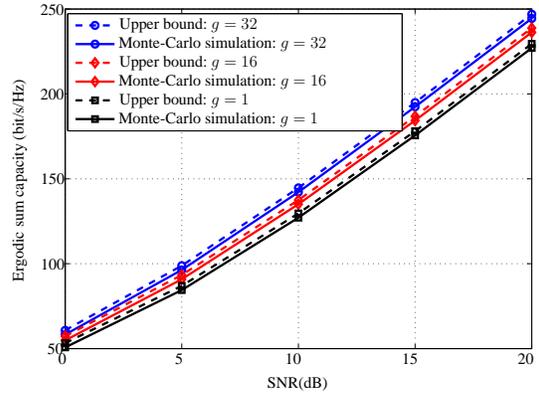}}
\caption{Ergodic sum capacity as a function of $g$, with $M=S=s_p=128$, $K=32$ and $\rho_p =1$.}
\label{changeg}
\end{figure}
\psfrag{40}[][]{\Large {40}}
\psfrag{50}[][]{\Large {50}}
\psfrag{60}[][]{\Large {60}}
\psfrag{Upper bound M=Ns=64,g=16 blablablablablabla}{\Large {Upper bound: $M=N=64$, $g=16$}}
\psfrag{data2}{\Large {Monte-Carlo simulation: $M=N=64$, $g=16$}}
\psfrag{Upper bound M=Ns=32,g=8}{\Large{Upper bound: $M=N=32$, $g=8$}}
\psfrag{data4}{\Large {Monte-Carlo simulation: $M=Ns=32$, $g=8$}}
\psfrag{Upper bound M=Ns=16,g=4}{\Large {Upper bound: $M=N=16$, $g=4$}}
\psfrag{data6}{\Large {Monte-Carlo simulation: $M=N=16$, $g=4$}}
\psfrag{Upper bound M=Ns=16,g=1}{\Large {Upper bound: $M=N=16$, $g=1$}}
\psfrag{data8}{\Large {Monte-Carlo simulation: $M=N=16$, $g=1$}}
\psfrag{70}[][l]{\Large {70}}
\psfrag{600}[][]{\Large {600}}
\psfrag{800}[][]{\Large {800}}
\psfrag{PV channel with g=M/5 aaaaaaaaaa}{\Large{A PV channel with $g=\frac{M}{5}$}}
\psfrag{WV channel with g=1}{\Large{A WV channel with $g=1$}}
\psfrag{M}[][cb]{\Large {Number of antennas}}
\begin{figure}[t]
\centering
\scalebox{0.43}{\includegraphics*{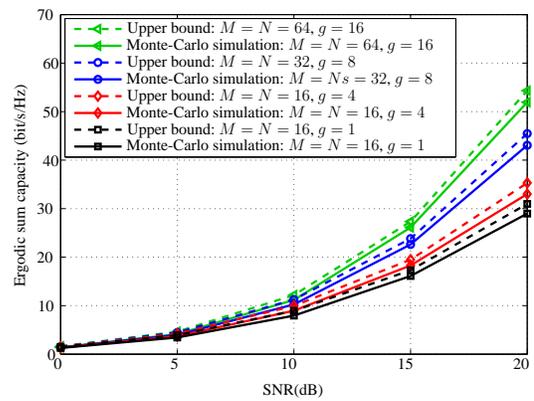}}
\caption{Ergodic sum capacity as as a function of $M$, $S$ and $g$, with $K =32$ and transmit energy reduced with $\frac{1}{M}$.}
\label{changeall}
\end{figure}

The influence of the cluster VR size on the sum capacity is shown in Fig.~\ref{changeg}. A notable increase of capacity is brought by increasing $g$ from 16 to 32 (i.e., the size of cluster VRs drop from 8 to 4), which justifies Proposition~\ref{corollary:2.1}. The spatially WSS channel (i.e., $g=1$) is also provided as a baseline.

Now we investigate the influence of size of antenna array by increasing $M$, $g$ and $S$ simultaneously with fixed $K$. Since a larger $M$ provides a higher array gain, the transmit power is reduced as $\frac{1}{M}$. The simulation results are shown in Fig.~\ref{changeall}. As shown in the figure, even though we reduce the transmit power, a significant increase of capacity is still obtained, which comes from the more significant spatial non-stationarity characteristics. When SNR is 15dB, the capacity with $M =64$ and $g=16$ is 10 bit/s/Hz higher than that with $M =16$ and $g=4$. The spatially WSS channel (i.e., $g=1$) is also provided as a baseline. Fig.~\ref{changeall} indicates that in massive MIMO channels, a capacity gain can be obtained not only from a higher array gain, but also from a more non-WSS channel.

\begin{figure}[t]
\centering
\scalebox{0.43}{\includegraphics*{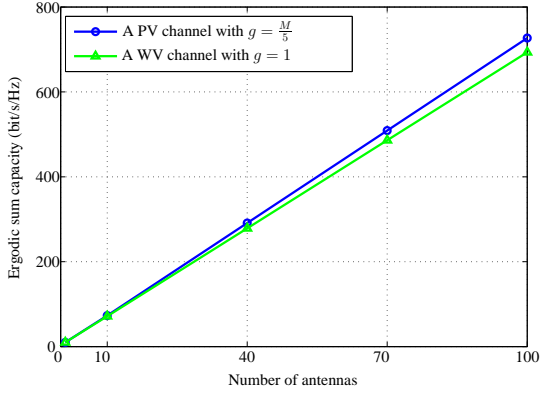}}
\caption{Ergodic sum capacity as as a function of $M=S=K$, with $g =\frac{M}{5}$ and $\tilde \mu_{\rm{dB}}=35\,{\rm {dB}}$.}
\label{changeall2}
\end{figure}

To see the asymptotic behavior of sum capacity in large system dimension, Fig.~\ref{changeall2} is further plotted with the total SNR $\tilde \mu_{\rm{dB}} = 10\log_{10}(\tilde \mu)=35\, {\rm{dB}}$ where $\tilde \mu$ is defined in Proposition~\ref{corollary:2.2}. The high SNR is selected so that the upper bound at high SNR is an accurate approximation of the real upper bound. As shown in the figure, the capacity increases linearly with $M$ when $M$ becomes large. The increasing rate of capacity is 7.1 bit/s/Hz for the complete PV channel, which is equal to that predicted by Proposition~\ref{corollary:2.2}. Moreover, it is shown that the complete PV channel has a higher increasing rate than the complete WV channel, and the capacity advantage becomes larger as $M$ increases.

\begin{figure}[!t]
\centering
\scalebox{0.43}{\includegraphics*{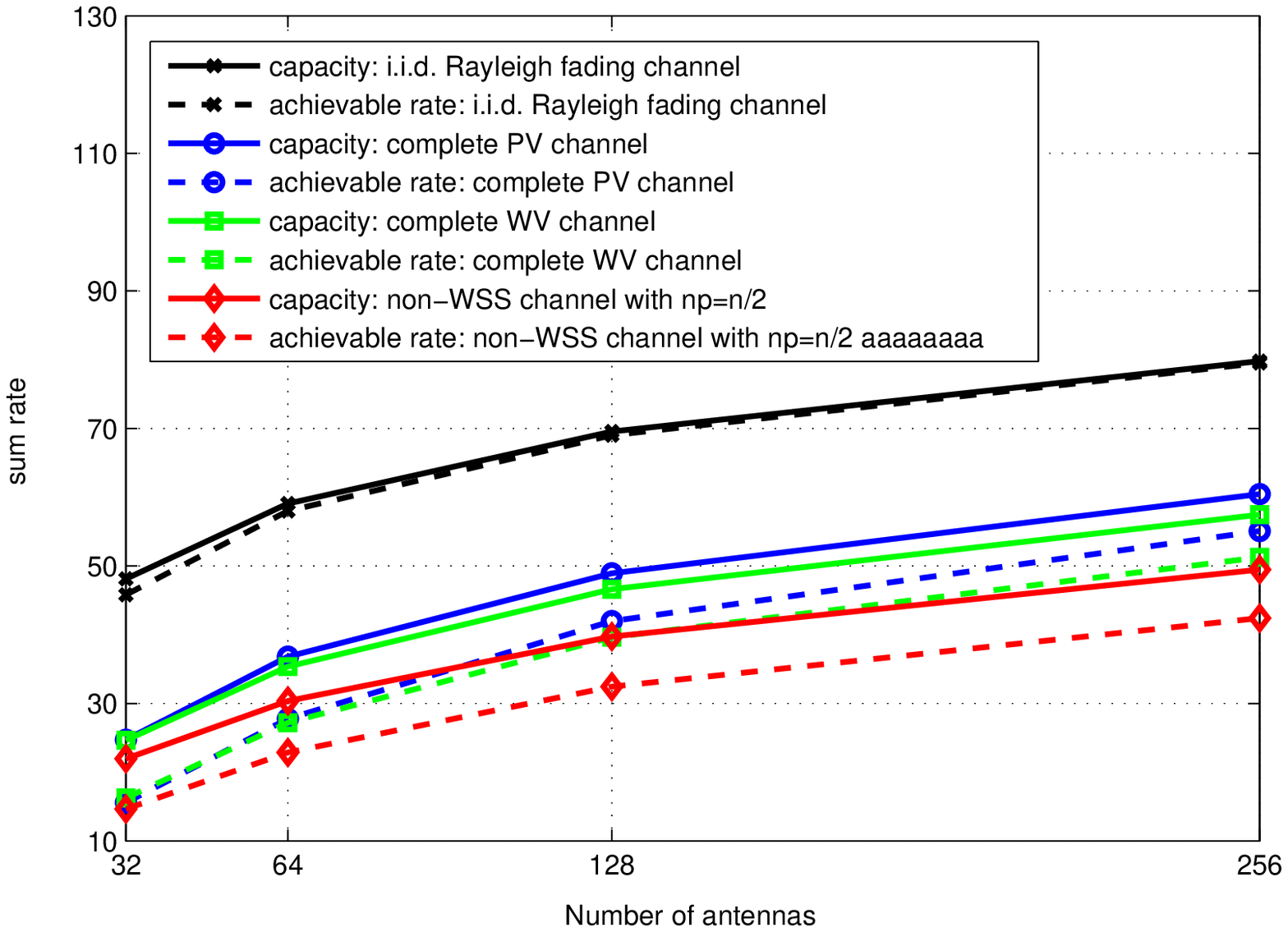}}
\caption{Ergodic sum capacity and achievable rate of the linear MMSE receiver in a i.i.d. Rayleigh fading channel, a complete PV channel, a channel with both PV and WV clusters, and a complete WV channel with $K=20$.}
\label{compare}
\end{figure}

Finally, we consider the performance gap between the sum capacity and the achievable rate achieved by linear MMSE receivers. In the i.i.d. Rayleigh fading channels, linear processing could achieve a sum rate that is close to the sum capacity in massive MIMO systems, which is one of the key motivations of massive MIMO systems in existing literature. Thus it inspires us to compare them in non-i.i.d. massive MIMO channels. The simulation results are provided in Fig.~\ref{compare}. As shown in the figure, for the i.i.d. Rayleigh fading channel, the performance gap between the sum capacity and the achievable rate is small and decreases rapidly as $M$ grows. However, in the other non-i.i.d. Rayleigh channels, the gaps are bigger and decrease very slowly as $M$ increases. It indicates that the use of the suboptimal linear MMSE receivers can cause notable performance losses in non-i.i.d. Rayleigh channels. Notice that although it is spatially WSS, the i.i.d. Rayleigh fading channel has no spatial correlation, and thus it has the most even spread of eigenvalues. Therefore, the i.i.d. fading channel gives the highest sum capacity. In non-i.i.d. channels, the complete PV channel provides the highest capacity, which is consistent with our conclusion that more significant spatially non-WSS characteristics benefit the sum capacity.
\psfrag{Sum capacity blabla}{\Large{Sum capacity}}
\psfrag{rate}{\Large{Achievable rate of MMSE receiver}}
\psfrag{Number of antennas}[][cb]{\Large {Number of antennas}}
\psfrag{sum rate}[][]{\Large{Sum capacity and achievable rate (bit/s/Hz)}}
\psfrag{10}[][]{\Large {10}}
\psfrag{30}[][]{\Large {30}}
\psfrag{50}[][]{\Large {50}}
\psfrag{70}[][]{\Large {70}}
\psfrag{90}[][]{\Large {90}}
\psfrag{110}[][]{\Large {110}}
\psfrag{130}[][]{\Large {130}}
\psfrag{32}[][l]{\Large {32}}
\psfrag{64}[][l]{\Large {64}}
\psfrag{128}[][l]{\Large {128}}
\psfrag{256}[][l]{\Large {256}}

\psfrag{capacity: i.i.d. Rayleigh fading channel}{\Large{Capacity: i.i.d. Rayleigh fading channel}}
\psfrag{achievable rate: i.i.d. Rayleigh fading channel}{\Large{Achievable rate: i.i.d. Rayleigh fading channel}}
\psfrag{capacity: complete PV channel}{\Large{Capacity: complete PV channel}}
\psfrag{achievable rate: complete PV channel}{\Large{Achievable rate: complete PV channel}}
\psfrag{capacity: complete WV channel}{\Large{Capacity: non-WSS channel with $s_p=\frac{S}{2}$}}
\psfrag{achievable rate: complete WV channel}{\Large{Achievable rate: non-WSS channel with $s_p=\frac{S}{2}$}}
\psfrag{capacity: non-WSS channel with np=n/2}{\Large{Capacity: complete WV channel}}
\psfrag{achievable rate: non-WSS channel with np=n/2 aaaaaaaa}{\Large{Achievable rate: complete WV channel}}

\section{Conclusions} \label{sec:conclusion}
In this paper, a new channel model is proposed for massive MIMO systems to describe the spatially non-WSS channel characteristics. Based on the new model, a closed-form expression of an upper bound on the ergodic sum capacity is further derived, and the influence of the spatial non-stationarity on sum capacity is analyzed. Analysis shows that in non-i.i.d. Rayleigh channels, the non-stationarity results from cluster partial visibility can benefit the sum capacity by bringing a more even spread of channel eigenvalues. This shows the advantage of using a large antenna array in a practical non-i.i.d. Rayleigh fading channel: the sum capacity will benefit not only from a higher array gain, but also a more spatially non-wide sense stationary channel. Simulation results validate our analysis and the tightness of the upper bound.

\appendices
\section{Proof of Theorem~\ref{theorem}}\label{sec:proof thrm1}
To derive the close form expression of $C_{up}$ in~(\ref{capacity:new}), we need to obtain the expression of $R{\rm{ }} \triangleq {\Eset}\left[ {\det \left( {{{\bf{I}}_{M}} + \mu {{\bf{G}}}{\bf{G}}^H} \right)} \right]$.
Define matrix
\begin{equation} \nonumber
{\bf{A}} = \left[ {\begin{array}{*{20}{c}}
{{\bf{W}}_p^{\frac{1}{2}}}&{{\bf{W}}_w^{\frac{1}{2}}}
\end{array}} \right]\left[ {\begin{array}{*{20}{c}}
{{{\bf{H}}_p}}&{\bf{0}}\\
{\bf{0}}&{{{\bf{H}}_w}}
\end{array}} \right],
\end{equation}
where ${{\bf{H}}_p} $ is defined in~(\ref{newmodel}), and the ${\bf W}_p$ and ${\bf W}_w$ are defined in~(\ref{eqn:correlation}), then ${{\bf{G}}}= {\bf{AR}}_s^{\frac{1}{2}}{{\bf{H}}}$. Inspired by~\cite{Shin2003capacity}, we apply the theorem of principal minor determinant expansion for the characteristic polynomial of a matrix in~\cite{Grant2002rayleigh}, and the Binet-Cauchy formula for the determinant of a product matrix in~\cite{Browne1958} to $R$ to obtain
\begin{eqnarray} \label{equ:R1}
R &=& \sum\limits_{k = 0}^{m} {{\left( \mu  \right)}^k}\sum\limits_{\scriptstyle1 \le {i_1} \le ..\hfill\atop
\scriptstyle \le {i_k} \le {S}\hfill} \sum\limits_{\scriptstyle1 \le {a_1} \le \ldots\hfill\atop
\scriptstyle \le {a_k} \le {S}\hfill} \sum\limits_{\scriptstyle1 \le {b_1} \le \ldots\hfill\atop
\scriptstyle \le {b_k} \le {S}\hfill}
\sum\limits_{\scriptstyle1 \le {c_1} \le \ldots\hfill\atop \scriptstyle \le {c_k} \le K\hfill}\nonumber \\
& &\sum\limits_{\scriptstyle1 \le {d_1} \le ..\hfill\atop
\scriptstyle \le {d_k} \le K\hfill}\left( \Eset \left[ {\det \left( {\left( {{{\bf{A}}^H}{\bf{A}}} \right)_{{a_1},\ldots,{a_k}}^{{i_1},\ldots,{i_k}}} \right)} \right]\right. \nonumber \\
&\times &{\det \left( {\left( {{\bf{R}}_s^{\frac{1}{2}}} \right)_{{b_1},\ldots,{b_k}}^{{a_1},\ldots,{a_k}}} \right)}  \det \left( {\left( {{\bf{R}}_s^{\frac{1}{2}}} \right)_{{i_1},\ldots,{i_k}}^{{d_1},\ldots,{d_k}}} \right) \nonumber \\
&\times &\left.\Eset \left[ {\det \left( {\left( {\bf{H}} \right)_{{c_1},\ldots,{c_k}}^{{b_1},\ldots,{b_k}}} \right)\det \left( {\left( {{{\bf{H}}^H}} \right)_{{d_1},\ldots,{d_k}}^{{c_1},\ldots,{c_k}}} \right)} \right]\right),
\end{eqnarray}
where $m=\min\{M, K, S \}$. According to Lemma \Rmnum{2}.1 in~\cite{Shin2003capacity},
\begin{eqnarray}
& \Eset \left[ {\det \left( {\left( {{{\bf{H}}}} \right)_{{c_1},\ldots,{c_k}}^{{b_1},\ldots,{b_k}}} \right)\det \left( {\left( {{\bf{H}}^H} \right)_{{d_1},\ldots,{d_k}}^{{c_1},\ldots,{c_k}}} \right)} \right]\nonumber \\
& = \left\{ \begin{array}{lc}
k!, & {\text{if}}\quad  {b_1} = {d_1},...,{b_k} = {d_k},\\ %
0,  & {\text{otherwise}}.
\end{array} \right.
\end{eqnarray}
Thus,~(\ref{equ:R1}) can be simplified as
\begin{eqnarray} \label{equ:R2}
R &=& \sum\limits_{k = 0}^m {{\left( \mu  \right)}^k}k! C^{k}_{K}\sum\limits_{\scriptstyle1 \le {i_1} \le \ldots\hfill\atop
\scriptstyle \le {i_k} \le {S}\hfill} \sum\limits_{\scriptstyle1 \le {a_1} \le \ldots\hfill\atop
\scriptstyle \le {a_k} \le {S}\hfill}  {\det \left( {\left( {{{\bf{R}}_s}} \right)_{{i_1},\ldots,{i_k}}^{{a_1},\ldots,{a_k}}} \right)}. \nonumber \\
& \times&\Eset \left[ {\det \left( {\left( {{{\bf{A}}^H}{\bf{A}}} \right)_{{a_1},\ldots,{a_k}}^{{i_1},\ldots,{i_k}}} \right)} \right].
\end{eqnarray}
Let ${\bf{B}} = [ {\begin{array}{*{20}{c}}
{{\bf{W}}_p^{\frac{1}{2}}}&{{\bf{W}}_w^{\frac{1}{2}}}
\end{array}} ] $ and ${\bf{D}} = {\rm{bdiag}\{{\bf{H}}_p ,{\bf{H}}_w \}}$, we get
\begin{equation}
{\bf{\bar A}}\triangleq {{\bf{A}}^H}{\bf{A}} = {{\bf{D}}^H}{{\bf{B}}^H}{\bf{BD}}.
\end{equation}
By applying the Binet-Cauchy formula for a product matrix again, we get
\begin{align*}
& \Eset\left[ {\det \left( {\left( {{{{{\bf{\bar A}}}}}} \right)_{{a_1},\ldots,{a_k}}^{{i_1},\ldots,{i_k}}} \right)} \right] \nonumber \\
& = \sum\limits_{\scriptstyle1 \le {u_1} \le \ldots\hfill\atop
\scriptstyle \le {u_k} \le 2M\hfill} {\sum\limits_{\scriptstyle \le {v_1} \le \ldots\hfill\atop
\scriptstyle \le {v_k} \le 2M\hfill} {\det \left( {\left( {{{\bf{B}}^H}{\bf{B}}} \right)_{{v_1},\ldots,{v_k}}^{{u_1},\ldots,{u_k}}} \right)} } \nonumber \\
& \times  \Eset\left[ {\det \left( {\left( {{{\bf{D}}^H}} \right)_{{u_1},\ldots,{u_k}}^{{i_1},\ldots,{i_k}}} \right)\det \left( {\left( {\bf{D}} \right)_{{a_1},\ldots,{a_k}}^{{v_1},\ldots,{v_k}}} \right)} \right]. \numberthis
\end{align*}
According to the definition of matrix determinant,
\begin{equation} \label{determinant}
\det \left( {\left( {\bf{D}} \right)_{{a_1},\ldots,{a_k}}^{{v_1},\ldots,{v_k}}} \right) = \sum\limits_{{\bf{p}}} {{{\left( { - 1} \right)}^{\tau \left( {\bf{p}} \right)}}{d_{{v_1}{p_1}}}{d_{{v_2}{p_2}}}\ldots{d_{{v_k}{p_k}}}} ,
\end{equation}
where ${\bf{p}} =( {{p_1},..,{p_k}} )$ represents a permutation of $( {{a_1}\ldots,{a_k}} )$, and $\tau ( {\bf{p}} )$ denotes the inverse number of $\bf{p}$. Then we have
\begin{align*}\label{constraint_D}
&\Eset\left[ {\det \left( {\left( {{{\bf{D}}^H}} \right)_{{u_1},\ldots,{u_k}}^{{i_1},\ldots,{i_k}}} \right)\det \left( {\left( {\bf{D}} \right)_{{a_1},\ldots,{a_k}}^{{v_1},\ldots,{v_k}}} \right)} \right] \nonumber \\
&= \sum\limits_{\bf{p}} {\sum\limits_{\bf{q}} {{{\left( { - 1} \right)}^{\tau \left( {\bf{p}} \right) + \tau \left( {\bf{q}} \right)}}\Eset\left[ {{d_{{v_1}{p_1}}}\ldots{d_{{v_k}{p_k}}}d_{{u_1}{q_1}}^*\ldots d_{{u_k}{q_k}}^*} \right]}} \numberthis.
\end{align*}
Due to the special structure of $\bf{D}$, ${{d_{{v_i}{p_i}}}}(i\in \{1,\ldots,k \})$ is either zero or an i.i.d. Gaussian variable with zero mean and unit variance. Consequently, the necessary condition for
\begin{equation}
\Eset\left[ {\det \left( {\left( {{{\bf{D}}^H}} \right)_{{u_1},\ldots,{u_k}}^{{i_1},\ldots,{i_k}}} \right)\det \left( {\left( {\bf{D}} \right)_{{i_1},\ldots,{i_k}}^{{u_1},\ldots,{u_k}}} \right)} \right] \ne 0
\end{equation}
is that
\begin{subequations}
\begin{numcases}{}
{u_1} = {v_1},\ldots,{u_k} = {v_k}, \\
{a_1} = {i_1},\ldots,{a_k} = {i_k}, \\
{p_1} = {q_1},\ldots,{p_k} = {q_k}.  \label{constraint_pq}
\end{numcases}
\end{subequations}
Then $R$ in~(\ref{equ:R2}) can be reduced as
\begin{eqnarray} \label{equ:R3}
R &=& \sum\limits_{k = 0}^m {{\left( \mu  \right)}^k}k! C^{k}_{K}\sum\limits_{\scriptstyle1 \le {i_1} \le \ldots\hfill\atop
\scriptstyle \le {i_k} \le {S}\hfill} \left(\det \left( {\left( {{{\bf{R}}_s}} \right)_{{i_1},\ldots,{i_k}}^{{i_1},\ldots,{i_k}}} \right)\right. \nonumber \\
&\times &\left. \Eset\left[ {\det \left( {\left( {{{\bf{A}}^H}{\bf{A}}} \right)_{{i_1},\ldots,{i_k}}^{{i_1},\ldots,{i_k}}} \right)} \right] \right),
\end{eqnarray}
and
\begin{align*}
&\Eset\left[ {\det \left( {\left( {{{\bf{A}}^H}{\bf{A}}} \right)_{{i_1},\ldots,{i_k}}^{{i_1},\ldots,{i_k}}} \right)} \right] = \sum\limits_{\scriptstyle1 \le {j_1} \le \ldots\hfill\atop
\scriptstyle \le {j_k} \le 2M \hfill}\left(\det \left( {\left( {{{{\bf{ R}}}_r}} \right)_{{j_1},\ldots,{j_k}}^{{j_1},\ldots,{j_k}}} \right) \right.\nonumber \\
&\times \left. \Eset\left[ {\det \left( {\left( {\bf{D}} \right)_{{i_1},\ldots,{i_k}}^{{j_1},\ldots,{j_k}}} \right)\det \left( {\left( {{{\bf{D}}^H}} \right)_{{j_1},\ldots,{j_k}}^{{i_1},\ldots,{i_k}}} \right)} \right]\right) \numberthis,
\end{align*}
where ${{\bf{R}}_r} = {{\bf{B}}^H}{\bf{B}}$. Define
\begin{align*}
&N\left( {{j_1},\ldots{j_k},{i_1},\ldots,{i_k}} \right)  \nonumber \\
&\triangleq \Eset\left[ {\det \left( {\left( {{{\bf{D}}^H}} \right)_{{j_1},\ldots,{j_k}}^{{i_1},\ldots,{i_k}}} \right)\det \left( {\left( {\bf{D}} \right)_{{i_1},\ldots,{i_k}}^{{j_1},\ldots,{j_k}}} \right)} \right]\numberthis,
\end{align*}
then from constraint~(\ref{constraint_pq}), we know that
\begin{equation}
N\left( {{j_1},\ldots,{j_k},{i_1},\ldots,{i_k}} \right) = \sum\limits_{\bf{p}} {\Eset\left( {{{\left| {{d_{{j_1}{p_1}}}} \right|}^2}} \right)\ldots \Eset\left( {{{\left| {{d_{{j_k}{p_k}}}} \right|}^2}} \right)}.
\end{equation}
Since ${d_{{j_n}{p_n}}} = 0$ or ${d_{{j_n}{p_n}}} \sim CN( {0,1} )$, thus for a particular $\bf{p}$, $N( {{j_1},\ldots,{j_k},{i_1},\ldots,{i_k}}) \ne 0$ leads to ${d_{{j_n}{p_n}}} \sim \mathcal{CN}( {0,1} )$ for all ${n \in \{ {1,\ldots,k} \}}$. Consequently, $N( {{j_1},\ldots,{j_k},{i_1},\ldots,{i_k}} )$ represents how many permutations of $( {{i_1},\ldots,{i_k}} )$, denoted as ${\bf{p}} = ( {{p_1},\ldots,{p_k}})$, can guarantee that ${d_{{j_n}{p_n}}} \sim \mathcal{CN}( {0,1} )$ for all ${n\in \{ {1,\ldots,k} \}}$. Assume that for a particular $( {{j_1},\ldots,{j_k},{i_1},\ldots,{i_k}} )$, $m_v$ rows and $n_v$ columns are selected from the $v$th diagonal block of ${\bf{D}}$ ($ v \in  \{1,\ldots,g+1 \}$, and we denote $r_{g+1}=M$ and $s_{g+1} = s_w$). Then
\begin{equation} \label{N value}
\sum\limits_{v = 1}^{g + 1} {{m_v}} = k, \quad \quad \sum\limits_{v = 1}^{g + 1} {{n_v}}  = k.
\end{equation}
Therefore, $N({{j_1},\ldots,{j_k},{i_1},\ldots,{i_k}}) \ne 0$ indicates that ${m_v} \le {n_v}$. Together with~(\ref{N value}),  we have
\begin{equation} \label{N value2}
{m_v} = {n_v}
\end{equation}
for all $v \in \{1,\ldots,g+1 \}$, which means $m_v$ rows and columns are selected from the $v$th diagonal block. Consequently, the number of permutations, $\bf{p}$, is $\prod\limits_{v = 1}^{g + 1} {{m_v}!}$. Thus,
\begin{align*}
    & N\left( {{j_1},\ldots,{j_k},{i_1},\ldots,{i_k}} \right) \nonumber \\
    & = \left\{ \begin{array}{lc}
    \prod\limits_{v = 1}^{g + 1} {{m_v}!}, & {\text{if}} \quad m_1 = n_1, \ldots, m_{g+1}=n_{g+1}, \\
    0, & {\text{otherwise}}. \numberthis
    \end{array} \right.
\end{align*}

Then $R$ in~(\ref{equ:R3}) is
\begin{eqnarray}
R &=& \sum\limits_{k = 0}^m {{\left( \mu  \right)}^k}k! C^{k}_{K}\sum\limits_{\scriptstyle1 \le {i_1} \le \ldots\hfill\atop
\scriptstyle \le {i_k} \le {S}\hfill} {\sum\limits_{\scriptstyle1 \le {j_1} \le \ldots \hfill\atop
\scriptstyle \le {j_k} \le 2M\hfill} {\det \left( {\left( {{{\bf{R}}_s}} \right)_{{i_1},\ldots,{i_k}}^{{i_1},\ldots,{i_k}}} \right)} } \nonumber \\
&\times& \det \left( {\left( {{{\bf{R}}_r}} \right)_{{j_1},\ldots,{j_k}}^{{j_1},\ldots,{j_k}}} \right) N\left( {{j_1},\ldots,{j_k},{i_1},\ldots,{i_k}} \right),
\end{eqnarray}
which completes the proof of Theorem~\ref{theorem}.

\section{Proof of Proposition~\ref{prop1}}\label{sec:proof prop1}
Under the conditions of Proposition~\ref{prop1}, we have at high SNR
\begin{align*}
 &{C_{up}} \approx n{\log _2}\mu  + {\log _2}\left( {n!} \right) + {\log _2}\det \left( {{{\bf{R}}_s}} \right) \nonumber \\
 & +  {\log _2}\sum\limits_{\scriptstyle1 \le {u_1} \le \ldots\hfill\atop
\scriptstyle \le {u_n} \le {2n}\hfill}\det \left( {\left( {{{{\bf{ R}}}_r}} \right)_{{u_1},\ldots,{u_n}}^{{u_1},\ldots,{u_n}}} \right) N\left( {{u_1},\ldots,{u_n},1,\ldots,n} \right),\numberthis
\end{align*}
where only the last term will change as $\rho_i$ changes.

Assume that $( {{u_1},\ldots,{u_n}} )$ selects ${m_v}$ rows from the $v$th block of ${\bf D}$. Since $(i_1,\ldots,i_n)=(1,\ldots,n )$ selects $s_{p,0}$ columns in the first $g$ blocks, and $s_w$ columns in the $(g+1)$th block, then similar to the analysis in Appendix~\ref{sec:proof thrm1}, we have
\begin{equation} \label{ms}
{m_v} = {s_{p,0}}, {m_{{g} + 1}} = {s_w},
\end{equation}
if $ N( {{u_1},\ldots,{u_n},1,\ldots,n} ) \ne 0$. Define ${\cal A}$ to be the set that contains all $( {{u_1},\ldots,{u_n}} )$ that satisfy condition A, then $\forall ( {{u_1},\ldots,{u_n}} ) \in {\cal A}$, $N( {{u_1},\ldots,{u_n}}, 1,\ldots,n ) = {( {{s_{p,0}}!} )^{{g}}}( {{s_w}!} )$. Consequently,
\begin{align*}
&\sum\limits_{\scriptstyle1 \le {u_1} \le \ldots\hfill\atop
\scriptstyle \le {u_n} \le 2n\hfill} \det \left( {\left( {{{{\bf{R}}}_r}} \right)_{{u_1},\ldots,{u_n}}^{{u_1},\ldots,{u_n}}} \right)N\left( {{u_1},\ldots,{u_n},1,\ldots,n} \right) \nonumber \\
& =  {\left( {{s_{p,0}}!} \right)^{{g}}}\left( {{s_w}!} \right)\sum\limits_{\left( {{u_1},\ldots,{u_n}} \right) \in {\cal A}} {\det \left( {\left( {{{{\bf{R}}}_r}} \right)_{{u_1},\ldots,{u_n}}^{{u_1},\ldots,{u_n}}} \right)} \numberthis.
\end{align*}
Define
\begin{equation}
{\bf{\tilde R}}_r = \left[ {\begin{array}{*{20}{c}}
{\bf R}_p & {{\bf R}_p^{\frac{1}{2}}{\bf{R}}_w^{\frac{1}{2}}}\\
{{\bf{R}}_w^{\frac{1}{2}}{\bf{R}}_p^{\frac{1}{2}}}&{{{{\bf{R}}}_w}}
\end{array}} \right],
\end{equation}
then according to~(\ref{ms}),
\begin{equation}
\left( {{{{\bf{R}}}_r}} \right)_{{u_1},\ldots,{u_n}}^{{u_1},\ldots,{u_n}} = \left[ {\begin{array}{*{20}{c}}
{\frac{{{\rho_p}}}{{{s_{p,0}}}}{\bf{A}}\left( {\bf{u}} \right)}&{\sqrt {\frac{{{\rho_p}\left( {1 - {\rho_p}} \right)}}{{{s_{p,0}}{s_w}}}} {\bf{X}}\left( {\bf{u}} \right)}\\
{\sqrt {\frac{{{\rho_p}\left( {1 - {\rho_p}} \right)}}{{{s_{p,0}}{s_w}}}} {{\bf{X}}^H}\left( {\bf{u}} \right)}&{\frac{{1 - {\rho_p}}}{{{s_w}}}{\bf{B}}\left( {\bf{u}} \right)}
\end{array}} \right],
\end{equation}
where ${\bf u}=(u_1,\ldots,u_n)$. ${\bf{A}}({\bf{u}}) $, ${\bf{X}}( {\bf{u}} )$ and ${\bf{B}}( {\bf{u}} )$ are submatrices extracted from ${{\bf R}_p}$, ${\bf R}_p^{\frac{1}{2}}{\bf R}_w^{\frac{1}{2}}$ and ${\bf R}_w$ according to ${\bf{u}} = ( {{u_1},\ldots,{u_n}} )$, respectively. Assume ${\bf{A}}({\bf{u}}) $ is invertible, then
\begin{align*}
&\det \left( {\left( {{{{\bf{ R}}}_r}} \right)_{{u_1},\ldots,{u_n}}^{{u_1},\ldots,{u_n}}} \right)\nonumber \\
&= {\left( {\frac{{{\rho_p}}}{{{s_{p,0}}}}} \right)^{{s_p}}}{\left( {\frac{{1 - {\rho_p}}}{{{s_w}}}} \right)^{{s_w}}}\det \left( {{\bf{A}}\left( {\bf{u}} \right)} \right)\det \left( {\bf Y}\left({\bf u}\right) \right) \nonumber \\
&= {\left( {\frac{{{\rho_p}}}{{{s_{p,0}}}}} \right)^{{s_p}}}{\left( {\frac{{1 - {\rho_p}}}{{{s_w}}}} \right)^{{s_w}}} {\det \left( {\left( {\bf{\tilde R}}_r \right)_{{u_1},\ldots,{u_n}}^{{u_1},\ldots,{u_n}}} \right)} \nonumber \\
&\triangleq  f\left( {{\rho_p}} \right) \numberthis,
\end{align*}
where ${\bf Y}({\bf u})={\bf B}( {\bf u}) - {\bf X}^H( {\bf u} ){\bf A}( {\bf u})^{-1}{\bf X}({\bf u})$.

Since ${\bf{\tilde R}}_r$ is a positive semi-definite matrix, each principal minor of ${\bf{\tilde R}}_r$ is non-negative. Thus,
\begin{equation}
\sum\limits_{\left( {{u_1},\ldots,{u_n}} \right) \in {\cal A}} {\det \left( {\left( {\bf{\tilde R}}_r \right)_{{u_1},\ldots,{u_n}}^{{u_1},\ldots,{u_n}}} \right)}  \ge 0.
\end{equation}
Define $l( x )=x^{s_p}( {1 - x} )^{s_w}( {x \in [ {0,1} ]} )$, then it is easy to prove that $l( x )$ is an increasing function of $x$ when $0 \le x \le \frac{s_p}{n}$, and a decreasing function when $\frac{s_p}{n} < x \le 1$. Thus,
\begin{equation}
f\left( {{\rho_p}} \right) = \frac{{l\left( {{\rho_p}} \right)}}{{{s_{p,0}}^{{s_p}}{s_w}^{{s_w}}}}\sum\limits_{\left( {{u_1},\ldots,{u_n}} \right) \in A} {\det \left( {\left( {\bf{\tilde R}}_r \right)_{{u_1},\ldots,{u_n}}^{{u_1},\ldots,{u_n}}} \right)}
\end{equation}
is an increasing function of $\rho_p$ when $0 \le \rho_p \le \frac{{{s_p}}}{n}$, and a decreasing function when $\frac{{{s_p}}}{n} < \rho_p \le 1$, which completes the proof of Proposition~\ref{prop1}.

\section{Proof of Corollary~\ref{corollary1.1}} \label{sec:proof corollary1.1}
When $s=s_p$, we have
\begin{equation}\label{determinant2}
\sum\limits_{\left( {{u_1},\ldots,{u_n}} \right) \in {\cal A}} {\det \left( {\left( {\bf{\tilde R}}_r \right)_{{u_1},\ldots,{u_n}}^{{u_1},\ldots,{u_n}}} \right)}  = \det {\left( {{{\bf{R}}_p}} \right)^g}.
\end{equation}
It is trivial to prove that the determinant of the complex Toeplitz matrix ${{\bf{\Omega }}_d}( a )$ is $\det ( {{{\bf{\Omega }}_d}( a )} ) = (1 - |a|^2)^{d - 1}$.
Define
\begin{equation}
f\left( g \right) = {\log _2}{\left( {\frac{n}{g}!} \right)^g} + {\log _2}{\left( {\frac{g}{n}} \right)^n} + {\log _2}\left\{\left(1 - |a|^2\right)^{n - g} \right\}.
\end{equation}
Then Corollary~\ref{corollary1.1} is proved if $f( g )$ is a monotone increasing function of $g$.

Let ${f_1}( g ) \triangleq {\log _2}{( {\frac{n}{g}!} )^g}$, then ${f_1}( g )$ is not differentiable since the factorial is only defined for non-negative integers. To facilitate analysis, a tight approximation of the factorial, Stirling's approximation, is applied. According to Stirling's approximation,
\begin{equation}
{\left( {\frac{n}{g}!} \right)^g} \approx {\left( {2\pi n} \right)^{\frac{g}{2}}}{g^{ - \left( {n + \frac{g}{2}} \right)}}{\left( {\frac{n}{e}} \right)^n}.
\end{equation}
Thus the derivative of ${f_1}( g )$ is
\begin{equation}
\frac{{\partial {f_1}}}{{\partial g}} = \frac{1}{{\ln 2}}\left( {\frac{1}{2}\ln \frac{{2\pi n}}{{eg}} - \frac{n}{g}} \right).
\end{equation}
Let ${f_2}( g )\triangleq {\log _2}{( {\frac{g}{n}} )^n}$ and ${f_3}( g )\triangleq {\log _2}\{(1 - |a|^2)^{n - g} \}$, then
\begin{equation}
\frac{{\partial {f_2}}}{{\partial g}} = \frac{1}{{\ln 2}}\frac{n}{g}, \quad \frac{{\partial {f_3}}}{{\partial g}} = -{\log_2}\left(1-|a|^2\right).
\end{equation}
Thus we get
\begin{eqnarray}
\frac{{\partial \left( f \right)}}{{\partial g}} &\triangleq& \frac{{\partial \left( {{f_1} + {f_2} + {f_3}} \right)}}{{\partial g}} \nonumber \\
&=& \frac{1}{{2\ln 2}}{\ln \frac{2\pi n}{eg}} +  {\log_2}\left(\frac{1}{1-|a|^2} \right).
\end{eqnarray}

In real environments, the number of PV cluster group is less than the number of BS antennas, therefore $g \in (1,n]$. As a result, ${\ln \frac{2\pi n}{eg}} >0$. Moreover, it is obvious that ${\log_2}(\frac{1}{1-|a|^2} )$ when $0<|a|<1$. Thus, it can be concluded that $f( g )$ is a monotonic increasing function of $g$ within the range of $g \in [ {1,n} ]$, and consequently,
\begin{equation}
f\left( {1} \right) < f\left( {g} \right),\quad \text{if} \quad g>1.
\end{equation}
Since $g>1$ holds in a complete PV environment, the following inequality holds and completes the proof of Corollary~\ref{corollary1.1}:
\begin{equation}
{\left. {{C_{max}}} \right|_{S = {s_p}}} - {\left. {{C_{max}}} \right|_{S = {s_w}}} = f\left( g \right) - f\left( 1 \right) > 0.
\end{equation}

\section{Proof of Proposition~\ref{corollary:2.2}} \label{sec:proof corollary2.2}
We have
\begin{eqnarray}
C_{up}&=& n{\log_2}{\frac{\tilde \mu}{n}} +{\log_2}\left(n! \right) + \log_2 {\rm {det}}\left({\bf R}_s\right) \nonumber \\
&+& \frac{n}{c} \log_2 {\rm {det}}\left(\frac{1}{c}{\bf R}_p \right) + \frac{n}{c} \log_2\left(c!\right),
\end{eqnarray}
where $\tilde \mu= \frac{P}{\sigma^2}$. Plugging the determinants of ${\bf R}_s = {\bf \Omega}_n(a_s)$ and ${\bf R}_p = {\bf \Omega}_{c}(a_p)$ in to~(\ref{c_up remark2.2}), and applying Stirling's approximation to $n!$, then we have
\begin{eqnarray}
\frac{{\partial C_{up}}}{{\partial n}} &= &{\log _2}\left( {\frac{\tilde \mu }{ec}} \right) + \frac{1}{2n \ln 2}\nonumber \\
 &+& \frac{1}{c}{\log _2}\left( {c!} \right) + \log_2 h\left(a_s, a_p\right),
\end{eqnarray}
where $h(a_s, a_p)= (1-|a_s|^2)(1-|a_p|^2)^{1-\frac{1}{c}}$. Since $h(a_s, a_p)>0$ for $|a_s|\in (0,1)$ and $|a_p|\in(0,1)$, then for any $|a_s|\in (0,1)$ and $|a_p|\in(0,1)$, there exists a $\mu_{0}$, such that when $\tilde \mu \ge \mu_0$, ${( {\frac{\tilde \mu }{ec}} )}h( a_s, a_p )\ge 1$. Then $\frac{{\partial C_{up}}}{{\partial n}} \ge 0$, and $C_{up}$ is an increasing function of $n$ for $ n\ge 1$. As $n$ goes to infinity, we further have
\begin{eqnarray}
\frac{{\partial C_{up}}}{{\partial n}} \to {\log _2}\left( {\frac{\tilde \mu }{ec}} \right) + \frac{1}{c}{\log _2}\left( {c!} \right) + \log_2 h\left(a_s, a_p\right),
\end{eqnarray}
which completes the proof of Proposition~\ref{corollary:2.2}.

\section*{Acknowledgment}
The authors would like to thank Hei Victor Cheng and Erik G. Larsson for their valuable suggestions during the preparation of this paper.

\bibliographystyle{IEEEtran}
\bibliography{refs_JP}

\begin{thebibliography}{10}
\providecommand{\url}[1]{#1}
\csname url@samestyle\endcsname
\providecommand{\newblock}{\relax}
\providecommand{\bibinfo}[2]{#2}
\providecommand{\BIBentrySTDinterwordspacing}{\spaceskip=0pt\relax}
\providecommand{\BIBentryALTinterwordstretchfactor}{4}
\providecommand{\BIBentryALTinterwordspacing}{\spaceskip=\fontdimen2\font plus
\BIBentryALTinterwordstretchfactor\fontdimen3\font minus
  \fontdimen4\font\relax}
\providecommand{\BIBforeignlanguage}[2]{{%
\expandafter\ifx\csname l@#1\endcsname\relax
\typeout{** WARNING: IEEEtran.bst: No hyphenation pattern has been}%
\typeout{** loaded for the language `#1'. Using the pattern for}%
\typeout{** the default language instead.}%
\else
\language=\csname l@#1\endcsname
\fi
#2}}
\providecommand{\BIBdecl}{\relax}
\BIBdecl

\bibitem{Marzetta10}
T.~L. Marzetta, ``Noncooperative cellular wireless with unlimited numbers of
  base station antennas,'' \emph{IEEE Trans. Wireless Communications}, vol.~9,
  no.~1, pp. 3590--3600, Nov. 2010.

\bibitem{Rusek13}
F.~Rusek, D.~Persson, K.~L. Buon, E.~G. Larsson, T.~L. Marzetta, O.~Edfors, and
  F.~Tufvesson, ``Scaling up {MIMO}: {Opportunities} and challenges with very
  large arrays,'' \emph{IEEE Trans. Signal Process.}, vol.~30, no.~1, pp.
  40--60, Jan. 2013.

\bibitem{Larsson14}
E.~G. Larsson, O.~Edfors, F.~Tufvesson, and T.~L. Marzetta, ``Massive {MIMO}
  for next generation wireless systems,'' \emph{IEEE Commun. Mag.}, vol.~52,
  no.~2, pp. 186--195, Feb. 2014.

\bibitem{bjornson2014massive}
E.~Bj{\"o}rnson, E.~G. Larsson, and M.~Debbah, ``Massive {MIMO} for maximal
  spectral efficiency: {How} many users and pilots should be allocated?''
  \emph{IEEE Trans. Wireless Commun}, Dec. 2014, submitted.

\bibitem{Yang13}
H.~Yang and T.~L. Marzetta, ``Performance of conjugate and zero-forcing
  beamforming in large-scale antenna systems,'' \emph{IEEE J. Sel. Areas
  Commun.}, vol.~31, no.~2, pp. 172--179, Feb. 2013.

\bibitem{Zhang13}
J.~W. Zhang, X.~J. Yuan, and P.~Li, ``Hermitian precoding for distributed
  {MIMO} systems with individual channel state information,'' \emph{IEEE J.
  Sel. Areas Commun.}, vol.~31, no.~2, pp. 241--250, Feb. 2013.

\bibitem{Hien13}
H.~Q. Ngo, E.~G. Larsson, and T.~L. Marzetta, ``Energy and spectral efficiency
  of very large multiuser {MIMO} systems,'' \emph{IEEE Trans. Commun.},
  vol.~61, no.~4, pp. 1436--1449, Apr. 2013.

\bibitem{bjornson2014optimal}
E.~Bj{\"o}rnson, L.~Sanguinetti, J.~Hoydis, and M.~Debbah, ``Optimal design of
  energy-efficient multi-user {MIMO} systems: {Is} massive {MIMO} the answer?''
  \emph{IEEE Trans. Wireless Commun}, to appear.

\bibitem{bjornson2014nonideal}
E.~Bj{\"o}rnson, M.~Matthaiou, and M.~Debbah, ``Massive {MIMO} with non-ideal
  arbitrary arrays: {Hardware} scaling laws and circuit-aware design,''
  \emph{IEEE Trans. Wireless Commun}, Jul. 2014, accepted.

\bibitem{Jose11}
J.~Jose, A.~Ashikhmin, T.~L. Marzetta, and S.~Vishwanath, ``Pilot contamination
  and precoding in multi-cell {TDD} systems,'' \emph{IEEE Trans. Wireless
  Commun.}, vol.~10, no.~8, pp. 2640--2651, Aug. 2011.

\bibitem{Zheng2014}
X.~Zheng, H.~Zhang, W.~Xu, and X.~You, ``Semi-orthogonal pilot design for
  massive {MIMO} systems using successive interference cancellation,'' in
  \emph{Proc. IEEE GLOBECOM}, Dec. 2014, pp. 3719--3724.

\bibitem{Wu2014}
X.~Wu and W.~Xu, ``Downlink performance analysis with enhanced multiplexing
  gain in multi-cell large-scale {MIMO} under pilot contamination,'' in
  \emph{Proc. IEEE WCNC}, Apr. 2014, pp. 230--235.

\bibitem{Shin2003}
H.~D. Shin and J.~H. Lee, ``Closed-form formulas for ergodic capacity of {MIMO
  Rayleigh} fading channels,'' in \emph{Proc. IEEE ICC}, vol.~5, May 2003, pp.
  2996--3000.

\bibitem{Hoydis2011asymptotic}
J.~Hoydis, R.~Couillet, and M.~Debbah, ``Asymptotic analysis of
  double-scattering channels,'' in \emph{Proc. Asilomar Conf.}, Nov. 2011, pp.
  1935--1939.

\bibitem{payami2012}
S.~Payami and F.~Tufvesson, ``Channel measurements and analysis for very large
  array systems at 2.6 {GHz},'' in \emph{Proc. 6th Eur. Conf. Antennas
  Propag.}, Mar. 2012, pp. 433--437.

\bibitem{Veerarajan2003}
T.~Veerarajan, \emph{Probability, Statistics and Random Processes}.\hskip 1em
  plus 0.5em minus 0.4em\relax Tata McGraw-Hill, 2003.

\bibitem{Yaghjian1986}
A.~D. Yaghjian, ``An overview of near-field antenna measurements,'' \emph{IEEE
  Trans. Antennas Propag.}, vol.~34, no.~1, pp. 30--45, Jan. 1986.

\bibitem{Gao13}
X.~Gao, F.~Tufvesson, and O.~Edfors, ``Massive {MIMO} channels - {Measurements}
  and models,'' in \emph{47th Annual Asilomar Conference on Signals, Systems,
  and Computers}, Nov. 2013, pp. 280--284.

\bibitem{Gao14}
X.~Gao, O.~Edfors, F.~Rusek, and F.~Tufvesson, ``Massive {MIMO} performance
  evaluation based on measured propagation data,'' \emph{IEEE Trans. Wireless
  Commun.}, vol.~14, no.~7, pp. 3899--3911, Jul. 2015.

\bibitem{Wu2014channel}
S.~Wu, C.~Wang, H.~Aggoune, M.~M. Alwakeel, and Y.~He, ``A non-stationary 3-{D}
  wideband twin-cluster model for {5G} massive {MIMO} channels,'' \emph{IEEE J.
  Sel. Areas Commun.}, vol.~32, no.~6, pp. 1207--1218, Jun. 2014.

\bibitem{correia2006mobile}
L.~M. Correia, \emph{{Mobile Broadband Multimedia Networks}}.\hskip 1em plus
  0.5em minus 0.4em\relax {Academic Press}, 2006.

\bibitem{liu2012cost}
L.~F. Liu, C.~Oestges, J.~Poutanen, K.~Haneda, P.~Vainikainen, F.~Quitin,
  F.~Tufvesson, and P.~D. Doncker, ``The {COST} 2100 {MIMO} channel model,''
  \emph{IEEE Wireless Commun.}, vol.~19, no.~6, pp. 92--99, 2012.

\bibitem{Huh2012}
H.~Huh, G.~Caire, H.~C. Papadopoulos, and S.~A. Ramprashad, ``Achieving
  `massive {MIMO}' spectral efficiency with a not-so-large number of
  antennas,'' \emph{IEEE Trans. Wireless Commun.}, vol.~11, no.~9, pp.
  3226--3239, Sept. 2012.

\bibitem{Yin2013}
H.~F. Yin, D.~Gesbert, M.~Filippou, and Y.~Z. Liu, ``A coordinated approach to
  channel estimation in large-scale multiple-antenna systems,'' \emph{IEEE J.
  Sel. Areas Commun.}, vol.~31, no.~2, pp. 264--273, Feb. 2013.

\bibitem{Gesbert02}
D.~Gesbert, H.~B\"{o}lcskei, D.~A. Gore, and A.~J. Paulraj, ``Outdoor {MIMO}
  wireless channels: {Models} and performance prediction,'' \emph{IEEE Trans.
  Commun.}, vol.~50, no.~12, pp. 1926--1934, Dec. 2002.

\bibitem{Ertel1998}
R.~B. Ertel, P.~Cardieri, K.~W. Sowerby, T.~S. Rappaport, and J.~H. Reed,
  ``Overview of spatial channel models for antenna array communication
  systems,'' \emph{IEEE Pers. Commun. Mag.}, pp. 10--22, Feb. 1998.

\bibitem{Fuhl1998}
J.~Fuhl, A.~F. Molisch, and E.~Bonek, ``Unified channel model for mobile radio
  systems with smart antennas,'' in \emph{Proc. IEE Radar, Sonar Navig.}, vol.
  145, Feb. 1998, pp. 32--41.

\bibitem{Zelst2002}
A.~V. Zelst and J.~S. Hammerschmidt, ``A single coefficient spatial correlation
  model for {MIMO} radio channels,'' in \emph{Proc. 27th General Assembly
  URSI}, 2002.

\bibitem{Chiani2003}
M.~Chiani, M.~Z. Win, and A.~Zanella, ``On the capacity of spatially correlated
  {MIMO} rayleigh-fading channels,'' \emph{IEEE Trans. Inform. Theory},
  vol.~49, no.~10, pp. 2363--2371, Oct. 2003.

\bibitem{Shin2006}
H.~Shin, M.~Z. Win, J.~H. Lee, and M.~Chiani, ``On the capacity of doubly
  correlated {MIMO} channels,'' \emph{IEEE Trans. Wireless Commun.}, vol.~5,
  no.~8, pp. 2253--2266, Aug. 2006.

\bibitem{Lozano2003}
A.~Lozano, A.~M. Tulino, and S.~Verd\'u, ``Multiple-antenna capacity in the
  low-power regime,'' \emph{IEEE Trans. Inform. Theory}, vol.~49, no.~10, pp.
  2527--2544, Oct. 2003.

\bibitem{Simon2000}
M.~K. Simon and M.-S. Alouin, \emph{Digital communication over fading channels:
  a unified approach to performance analysis}.\hskip 1em plus 0.5em minus
  0.4em\relax New York: Wiley, 2000.

\bibitem{David2005fundamentals}
D.~Tse and P.~Viswanath, \emph{{Fundamentals of Wireless
  Communications}}.\hskip 1em plus 0.5em minus 0.4em\relax {Cambridge
  University Press}, 2005.

\bibitem{Shin2003capacity}
H.~Shin and J.~H. Lee, ``Capacity of multiple-antenna fading channels: spatial
  fading correlation, double scattering, and keyhole,'' \emph{IEEE Trans. Inf.
  Theory}, vol.~49, no.~10, pp. 2636--2647, Oct. 2003.

\bibitem{Li2010}
X.~Li, S.~Jin, X.~Gao, and M.~R. McKay, ``Capacity bounds and low complexity
  transceiver design for double-scattering {MIMO} multiple access channels,''
  \emph{IEEE Trans. Signal Process.}, vol.~58, no.~5, pp. 2809--2822, May 2010.

\bibitem{Grant2002rayleigh}
A.~Grant, ``Rayleigh fading multi-antenna channels,'' \emph{{EURASIP J. Appl.
  Signal Processing (Special Issue on Space-Time Coding (Part I))}}, vol. 2002,
  no.~3, pp. 316--329, Mar. 2002.

\bibitem{Browne1958}
E.~T. Browne, \emph{{Introduction to the Theory of Determinants and
  Matrics}}.\hskip 1em plus 0.5em minus 0.4em\relax {Chapel Hill, NC: Univ.
  North Carolina Press}, 1958.

\end{thebibliography}

\end{document}